\documentclass[twocolumn,tighten]{aastex63}
\usepackage{graphicx,hyperref,amsmath,natbib}

\accepted{3 December 2019}
\submitjournal{AJ}

\shorttitle{Chemical Evolution of M31 dSphs}
\shortauthors{Kirby et al.}
\newcommand{\citeposs}[1]{\citeauthor{#1}'s (\citeyear{#1})}

\newcommand{\ncmdonly}{8}
\newcommand{\maxmratio}{120}

\newcommand{\maxmratioerrlo}{34}
\newcommand{\maxmratioerrhi}{48}
\newcommand{\maxmratiodsph}{And~V}
\newcommand{\minmratio}{14}

\newcommand{\minmratioerrlo}{9}
\newcommand{\minmratioerrhi}{24}
\newcommand{\minmratiodsph}{And~I}
\newcommand{\medfehdiff}{+0.25}
\newcommand{\meanfehdiff}{+0.30}
\newcommand{\meanfehdifferr}{0.02}

\newcommand{\stddevfehdiff}{0.32}
\newcommand{\stddevfehdifferr}{0.02}
\newcommand{\stddevfehdiffnorm}{1.53}
\newcommand{\stddevfehdiffnormerr}{0.11}
\newcommand{\stddevfehdiffnormcut}{1.37}
\newcommand{\stddevfehdiffnormcuterr}{0.10}
\newcommand{\medalphafediff}{-0.03}
\newcommand{\meanalphafediff}{+0.05}
\newcommand{\meanalphafedifferr}{0.05}

\newcommand{\stddevalphafediff}{0.37}
\newcommand{\stddevalphafedifferr}{0.03}
\newcommand{\stddevalphafediffnorm}{1.06}
\newcommand{\stddevalphafediffnormerr}{0.08}
\newcommand{\ntot}{256}
\newcommand{\nfehtot}{241}
\newcommand{\nalphafetot}{163}
\newcommand{\minsn}{4}
\newcommand{\maxsn}{73}
\newcommand{\medsn}{21}

\newcommand{\AndXfehmean}{-2.27}
\newcommand{\AndXfehmeanerr}{0.03}

\newcommand{\AndXnalphafe}{9}
\newcommand{\halofehmean}{-1.12}
\newcommand{\halofehmeanerr}{0.03}
\newcommand{\halomass}{8}
\newcommand{\halomasserr}{4}
\defcitealias{kir13b}{K13}
\defcitealias{var14a}{V14a}

\begin{document}

\title{Elemental Abundances in M31: The Kinematics and Chemical
  Evolution of Dwarf Spheroidal Satellite Galaxies\footnote{The data
    presented herein were obtained at the W.~M.~Keck Observatory,
    which is operated as a scientific partnership among the California
    Institute of Technology, the University of California and the
    National Aeronautics and Space Administration. The Observatory was
    made possible by the generous financial support of the W.~M.~Keck
    Foundation.}}

\correspondingauthor{Evan N. Kirby}
\email{enk@astro.caltech.edu}

\author[0000-0001-6196-5162]{Evan N. Kirby}
\affiliation{California Institute of Technology, 1200 E.\ California Blvd., MC 249-17, Pasadena, CA 91125, USA}

\author[0000-0003-0394-8377]{Karoline M. Gilbert}
\affiliation{Space Telescope Science Institute, 3700 San Martin Dr., Baltimore, MD 21218, USA} 
\affiliation{Department of Physics \& Astronomy, Bloomberg Center for Physics and Astronomy, Johns Hopkins University, 3400 N.\ Charles Street, Baltimore, MD 21218}

\author[0000-0002-9933-9551]{Ivanna Escala}
\affiliation{California Institute of Technology, 1200 E.\ California Blvd., MC 249-17, Pasadena, CA 91125, USA}
\affiliation{Department of Astrophysical Sciences, Princeton University, 4 Ivy Lane, Princeton, NJ 08544}

\author[0000-0002-3233-3032]{Jennifer Wojno}
\affiliation{Department of Physics \& Astronomy, Bloomberg Center for Physics and Astronomy, Johns Hopkins University, 3400 N.\ Charles Street, Baltimore, MD 21218}

\author[0000-0001-8867-4234]{Puragra Guhathakurta}
\affiliation{Department of Astronomy \& Astrophysics, University of California, Santa Cruz, 1156 High Street, Santa Cruz, CA 95064, USA}

\author[0000-0003-2025-3147]{Steven R. Majewski}
\affiliation{Department of Astronomy, University of Virginia, Charlottesville, VA 22904-4325, USA}

\author[0000-0002-1691-8217]{Rachael L. Beaton}
\altaffiliation{Hubble Fellow, Carnegie-Princeton Fellow}
\affiliation{Department of Astrophysical Sciences, Princeton University, 4 Ivy Lane, Princeton, NJ 08544}
\affiliation{The Observatories of the Carnegie Institution for Science, 813 Santa Barbara St., Pasadena, CA~91101}


\begin{abstract}

We present deep spectroscopy from Keck/DEIMOS of Andromeda I, III, V,
VII, and X, all of which are dwarf spheroidal satellites of M31.  The
sample includes \ntot\ spectroscopic members across all five dSphs.
We confirm previous measurements of the velocity dispersions and
dynamical masses, and we provide upper limits on bulk rotation.  Our
measurements confirm that M31 satellites obey the same relation
between stellar mass and stellar metallicity as Milky Way (MW)
satellites and other dwarf galaxies in the Local Group.  The
metallicity distributions show similar trends with stellar mass as MW
satellites, including evidence in massive satellites for external
influence, like pre-enrichment or gas accretion.  We present the first
measurements of individual element ratios, like [Si/Fe], in the M31
system, as well as measurements of the average [$\alpha$/Fe] ratio.
The trends of [$\alpha$/Fe] with [Fe/H] also follow the same galaxy
mass-dependent patterns as MW satellites.  Less massive galaxies have
more steeply declining slopes of [$\alpha$/Fe] that begin at lower
[Fe/H]\@.  Finally, we compare the chemical evolution of M31
satellites to M31's Giant Stellar Stream and smooth halo.  The
properties of the M31 system support the theoretical prediction that
the inner halo is composed primarily of massive galaxies that were
accreted early.  As a result, the inner halo exhibits higher [Fe/H]
and [$\alpha$/Fe] than surviving satellite galaxies.

\end{abstract}



\section{Introduction}
\label{sec:intro}

\subsection{Formation of Stellar Halos}

Dwarf galaxies have been established as the building blocks of stellar
halos.  Theoretical simulations \citep[e.g.,][]{bul05,joh08} predict
that more massive dwarf galaxies were accreted early in the formation
of an $L_*$ galaxy.  As a result, their stars populate the inner
portions ($r \lesssim 30$~kpc) of the halo.  Less massive dwarf
galaxies contribute later, sometimes dissolving into tidal streams.
Observational evidence in the Milky Way (MW) supports these
predictions.  Proper motions from the {\it Gaia} satellite have
discovered MW field stars that probably belonged originally to a
now-accreted galaxy, dubbed Gaia-Enceladus, with a stellar mass of
$\sim 6 \times 10^8~M_{\sun}$ \citep{bel18,hel18}.  The MW also shows
evidence of more recent accretion of smaller satellites, like the
Sagittarius dwarf spheroidal galaxy (dSph), which has a stellar mass
of $\sim 10^8~M_{\sun}$ \citep{nie10} and is embedded in a
well-defined tidal stream \citep{iba94,maj03}.

The elemental abundances of stars help disentangle the accretion
histories of stellar halos.  Detailed abundance ratios, like
[O/Fe]\footnote{We use the notation ${\rm [A/B]} = \log[n({\rm
      A})/n({\rm B})] - \log[n_{\sun}({\rm A})/n_{\sun}({\rm B})]$,
  where $n({\rm A})$ is the number density of atom A\@.  In this
  paper, we use the solar abundances of \citet{and89} except in the
  case of Fe, for which $\epsilon({\rm Fe}) = 12 + \log[n({\rm
      Fe})/n({\rm H})] = 7.52$.} and [Mg/Fe], are lower in metal-poor
(${\rm [Fe/H]} \lesssim -1$) stars in MW dSphs compared to the MW halo
\citep{she01,she03,ven04,hay18}.  This observation affirms that the
majority of the stellar mass of the MW halo was built of systems with
different chemical evolution histories from the presently surviving
dSph satellites.  Even the bulk metallicities of stars are informative
because dwarf galaxies obey a universal relationship between stellar
mass and stellar metallicity \citep[][hereafter
  \citetalias{kir13b}]{mat98,gre03,kir11a,kir13b}.  Therefore, the
metallicity distribution of the halo can be used to estimate the mass
function of accreted galaxies.  Observationally, the
metallicities---like the kinematics---suggest that the majority of
stars come from more massive dwarf galaxies \citep[$M_* \sim
  10^8~M_{\sun}$,][]{dea15,dea16,dlee15}.

While the MW has provided a trove of kinematical and abundance
information from photometry, spectroscopy, and astrometry, it is only
a single galaxy in a Universe filled with cosmic variance.  In fact,
the MW may be unusual in several regards.  For example, most MW-like
galaxies lack a satellite with the luminosity of the Large Magellanic
Cloud \citep[$1.5 \times 10^9~L_{\sun}$,][]{dev91}, and even fewer
have two Magellanic-type satellites \citep{tol11,bus11}.  Furthermore,
the Local Group is unusual for having so few star-forming satellites
\citep{geh17}.  There is also tentative evidence that the MW halo is
lower in stellar mass and more metal-poor than halos of mass similar
to the MW \citep{mon16,brauer19}.  We can achieve a more complete view
of galaxy assembly and evolution by examining the halo and satellites
of galaxies other than the MW\@.

The Great Andromeda Galaxy (M31) is an obvious alternative to the
MW\@.  The entire M31 system is confined to a portion of the sky, as
opposed to the MW, which occupies the entire sky.  At the same time,
its proximity \citep[785~kpc,][]{mcc05} still permits detailed
observations, including resolved stellar spectroscopy.  M31 also is a
good laboratory for hierarchical formation because it hosts dozens of
satellite galaxies \citep[i.a.,][]{mcc12}, and it has several
prominent tidal streams \citep{iba01,iba14,ric11}.

This paper is part of a series on the elemental abundances of
individual stars in the halo, Giant Stellar Stream (GSS), outer disk,
and dwarf satellite galaxies of M31.  \citet{esc19b} presented
detailed abundance ratios of 11 stars in M31's smooth halo.
\citet{gil19} showed similar measurements for 21 stars in the GSS,
along with a comparison to the abundances of dwarf galaxies.
\citet{esc19a} expanded the sample to 70 additional stars in M31's
smooth halo, kinematically cold substructure, and outer disk.  They
also performed a comparative analysis between the structural
components of M31, including a comparison of the inner stellar halo to
M31 dSphs.  This paper presents spectroscopy of \ntot\ member stars in
five M31 satellite galaxies.  The sample includes measurements of
[Fe/H] for \nfehtot\ stars and [$\alpha$/Fe] for \nalphafetot\ stars.
We discuss our spectroscopic observations and measurements in
Section~\ref{sec:spec}.

\subsection{Learning About Dwarf Galaxies from Resolved Stellar Spectroscopy}

In addition to its utility in probing the signatures of hierarchical
formation, our sample is also useful for studying the properties of
dwarf galaxies for their own sake.  For example, the radial velocities
of stars in dwarf galaxies can be used in complementarity with
metallicities to infer the masses and formation histories of those
galaxies.  (See \citealt{kal09,kal10}, \citealt{tol12}, and
\citealt{col13} for past kinematic studies of M31 dSphs.)  Radial
velocities can also reveal the presence of rotation or kinematic
substructure.  We discuss the kinematic properties of the M31 dSphs in
Section~\ref{sec:v}.

The metallicities of stars in a galaxy encode the galaxy's history of
nucleosynthesis, star formation, and gas flow \citep{tin80}.  Each
galactic component in which the metals are measured provides
complementary information.  The composition of the cold interstellar
medium reflects the metals available for new star formation.  Metals
in the warm circumgalactic medium have been expelled from past
episodes of gas outflow.  Finally, metals in stars record the history
of chemical evolution.  The stellar metallicity distribution function
(MDF) is the product of the star formation history (SFH) with the
age--metallicity relation: $dM_*/dZ_* = dM_*/dt \, \times \, dt/dZ_*$,
where $M_*$ represents the stellar mass at time $t$, and $Z_*$
represents the stellar metallicity.

Even the average stellar metallicity of the galaxy (the zeroth moment
of the MDF) is informative.  For gas-free galaxies, like the dSphs
under consideration here, the stellar MDF represents all of the metals
the galaxy has retained.  In other words, it is the difference between
all the metals created by stars and all of the metals lost due to
outflows or stripping.  To first order, metal creation is a function
of stellar mass.  (More stars will make more metals.)  Metal outflow
is a function of the gravitational potential.  (A deeper potential
allows the galaxy to retain more metals.)  This simple description
explains the mass--metallicity relation (MZR) for dwarf galaxies
(\citealt{leq79}; \citealt{ski89}; \citealt{kir11a};
\citetalias{kir13b}).  Galaxies with low $M_*$ also have low $Z_*$,
which indicates that they retained a smaller fraction of the metals
created by their stars than galaxies with high $M_*$ \citep{kir11c}.
We discuss the particular case of the metallicities of the M31 dSphs
in Section~\ref{sec:feh}.

In the beginning of a galaxy's life, the only type of supernova to
explode is core collapse (e.g., Type~II)\@.  Core collapse supernovae
produce a high ratio of the abundances of $\alpha$ elements, like O
and Mg, to iron.  As the galaxy progresses in its chemical evolution,
Type~Ia supernovae explode and produce copious amounts of iron but a
small amount of $\alpha$ elements.  As a result, the [$\alpha$/Fe]
ratio of a galaxy generally declines over time.  Bursts of star
formation can temporarily enhance [$\alpha$/Fe] because the
accompanying core collapse supernovae eject large amounts of $\alpha$
elements.  However, dwarf galaxies are particularly susceptible to a
steep decline in [$\alpha$/Fe] vs.\ [Fe/H]\@.  First, dwarf galaxies
generally have very low star formation rates (SFRs), as revealed
through direct observations of star-forming dwarfs \citep{lee09} and
measurements of the SFHs of quiescent dwarfs \citep{dol02,wei14}.
Second, their chemical evolution is slow.  For example, the rate at
which [Fe/H] increased in Local Group satellite galaxies was probably
slower than in the primary progenitors of the halos of the MW or M31
\citep{rob05,fon06,joh08}.

The chemical evolution of MW satellite galaxies has been documented
extensively
\citep[e.g.,][]{sun93,she98,she01,she03,gei05,kir09,kir11b,tol09,let10,lem14,hil19}.
In contrast, the measurements of detailed abundances of individual
stars outside of the MW's virial radius has been limited by the stars'
faintness.  So far, the only measurements of [$\alpha$/Fe] outside of
the MW's virial radius are for stars in M31's dSphs and halo
(\citealt{var14a}, hereafter \citetalias{var14a}; \citealt{var14b};
\citealt{gil19}; \citealt{esc19a,esc19b}) and stars in a few Local
Group dIrrs \citep{kir17a}.  This paper expands the body of detailed
abundance measurements in M31 dSphs, including measurements of
individual element ratios, such as [Si/Fe].  Section~\ref{sec:alphafe}
presents our contribution to the study of detailed abundances in M31
dSphs.

After we discuss the properties of the satellites themselves, we
return to their role in the assembly of M31's stellar halo in
Section~\ref{sec:discussion}.  Finally, we summarize the paper in
Section~\ref{sec:summary}.


\section{Spectroscopy}
\label{sec:spec}

We obtained spectroscopy with Keck/DEIMOS \citep{fab03} of individual
red giants in five dSphs: Andromeda VII, I, III, V, and X\@.  They are
numbered in the order in which they were discovered: And~I and III
\citep{van72}, And~V \citep{arm98}, And~VII \citep{kar99}, and And~X
\citep{zuc07}.  However, in this paper, we list them in decreasing
order of stellar mass whenever they appear in a list (e.g., in tables
and in figures).

\subsection{Target Selection}
\label{sec:target}

The deep spectroscopic slitmasks were designed using results from
slitmasks previously observed for approximately one hour as part of
the Spectroscopic Landscape of Andromeda's Stellar Halo survey
\citep[SPLASH, e.g.,][]{guh06,gil09,gil12}.  These masks, obtained by
\citet{kal09,kal10} and \citet{tol12}, were based on photometry
obtained by \citet[][And~VII]{gre99}, \citet[][And~I and III]{ost03},
\citet[][And~V]{bea14}, and \citet[][And~X]{zuc07}.  However, the
photometry used to design slitmasks presented in this paper is
exclusively from the Mosaic imager at Kitt Peak National Observatory.
The photometry was obtained by \citet{bea14} in the Washington system
($M$ and $T_2$ filters), which we transformed to the Cousins system
($V$ and $I$ filters) following \citet{maj00}.  Images were also
obtained with the DDO51 filter.  The DDO51 bandpass contains the Mg~b
triplet, which is a feature sensitive to surface gravity, and the
$M-{\rm DDO51}$ color then serves as a photometric proxy to
distinguish dwarfs and giants of the same $M-T_2$ color.  The previous
slitmasks (with shallower observations) used this diagnostic to select
against foreground dwarf stars, which have stronger Mg~b triplets than
red giant branch (RGB) stars.

\begin{figure*}
\centering
\includegraphics[width=\linewidth]{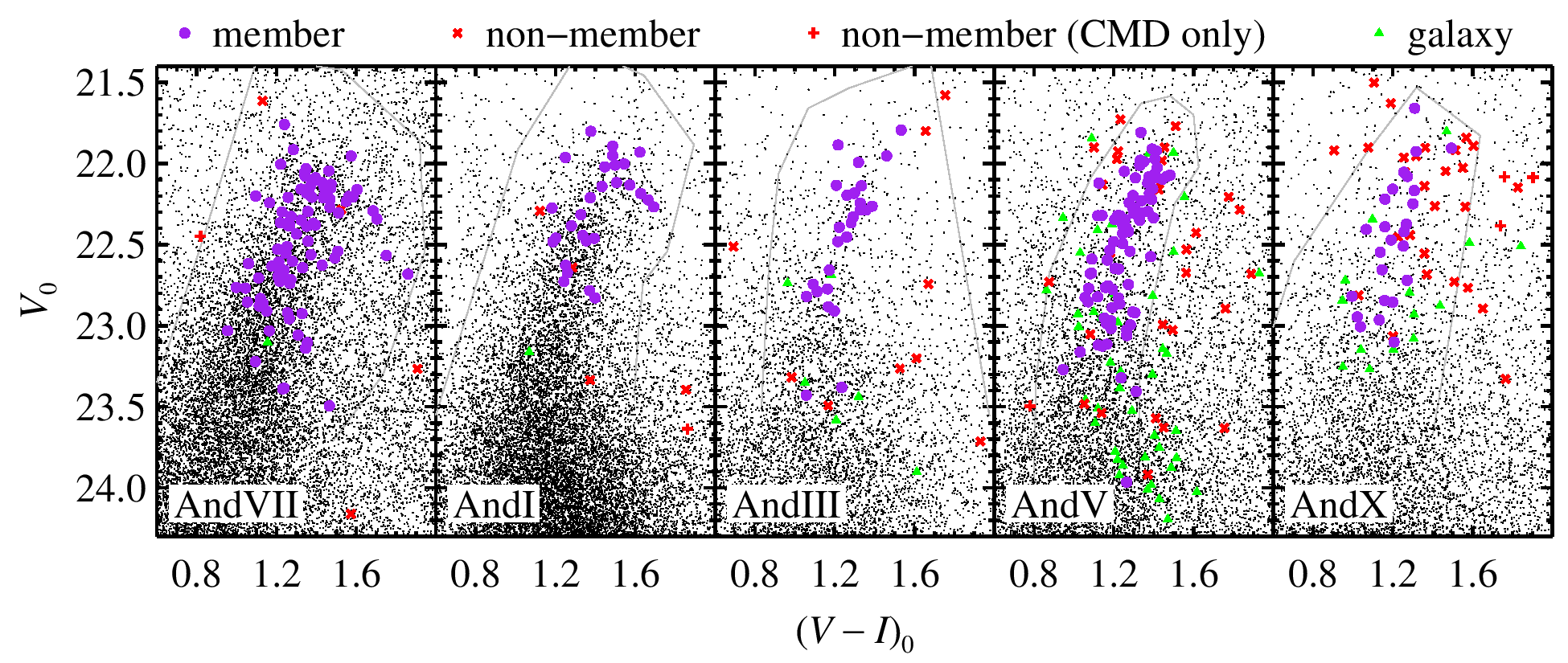}
\caption{CMDs of the M31 dSphs.  Colored points show DEIMOS
  spectroscopic targets.  Purple circles show spectroscopically
  confirmed member stars, as described in
  Section~\ref{sec:membership}.  Red $\times$'s indicate stars with a
  spectroscopic disqualifier for membership (radial velocity or a
  strong \ion{Na}{1} doublet).  Stars ruled as non-members on the
  basis of CMD position alone are represented by red $+$ signs.
  Unresolved galaxies are shown as green triangles.  The many small
  black points indicate objects without spectra.
  Section~\ref{sec:target} describes the spectroscopic target
  selection.  The gray polygons define the boundaries of the CMD
  membership criterion, discussed in
  Section~\ref{sec:membership}.\label{fig:cmd}}
\end{figure*}

As with most imaging spectrographs, the design of DEIMOS slitmasks
required us to choose some targets for spectroscopy at the expense of
others.  We developed a priority scheme to maximize the number of dSph
members.  Stars identified from previous slitmasks as likely to be RGB
stars at the distance of the dSph were included on the mask with
highest priority, while objects previously identified as MW foreground
stars or background galaxies were given lowest priority in the target
list.  High priority was also given to M31 dSph stars with existing
literature abundance measurements \citepalias{var14a}.  Additional
spectroscopic targets were chosen using the prioritization applied for
the SPLASH survey masks, which prioritized stars near the tip of the
RGB of each dSph over stars above or significantly below the tip.
Finally, we drew a wide polygon in the color--magnitude diagram (CMD)
around the RGB for each dSph, as shown in Figure~\ref{fig:cmd}.  Stars
outside of the polygons were given low priority for target selection.
The polygons were drawn so that any star that could reasonably belong
to the RGB was included.

\begin{figure*}
\centering
\includegraphics[width=\linewidth]{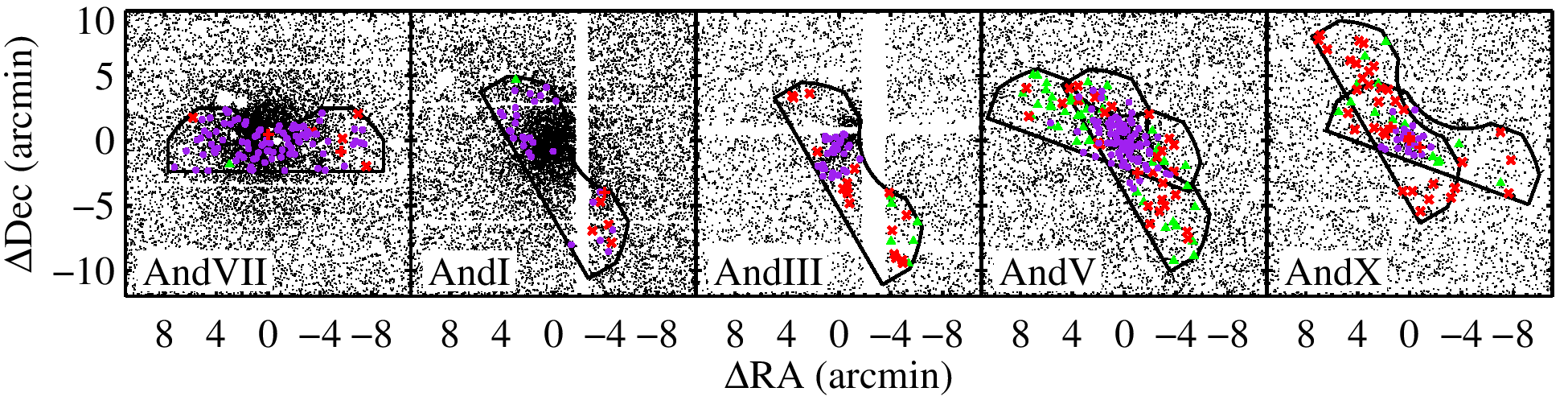}
\caption{Maps showing the positions and orientations of the DEIMOS
  slitmasks.  The irregular shapes delineate DEIMOS's field of view.
  Coordinates are relative to the galaxy centers \citep{tol12}.
  Symbols and colors have the same meaning as in Figure~\ref{fig:cmd}.
  Section~\ref{sec:target} describes the slitmask
  design.\label{fig:map}}
\end{figure*}

Figure~\ref{fig:cmd} shows the CMDs of the five dSphs.
Figure~\ref{fig:map} shows the orientation of the slitmasks on the
sky.  Stars confirmed to be spectroscopic members, as described in
Section~\ref{sec:membership}, are indicated by purple circles in both
figures.

\subsection{Observations and Data Reduction}
\label{sec:obs}

\begin{deluxetable*}{llcccccc }
\tablewidth{0pt}
\tablecolumns{8}
\tablecaption{DEIMOS Observations\label{tab:obs}}
\tablehead{\colhead{Galaxy} & \colhead{Slitmask} & \colhead{Slits} & \colhead{Tot.\ Exp.\ Time} & \colhead{Exp.\ Time} & \colhead{Exposures} & \colhead{Seeing} & \colhead{UT Date} \\
\colhead{ } & \colhead{ } & \colhead{ } & \colhead{(hr)} & \colhead{(min)} & \colhead{per visit} & \colhead{($''$)} & \colhead{ }}
\startdata
Andromeda~VII & and7a  &    119 & 4.6 & \phn71 & 3 & 0.9 & 2016 Sep 25    \\  
              &        &        &     & \phn53 & 3 & 0.8 & 2016 Sep 26    \\  
              &        &        &     & \phn80 & 3 & 0.9 & 2016 Sep 27    \\  
              &        &        &     & \phn72 & 3 & 1.1 & 2017 Jan 1\phn \\  
Andromeda~I   & and1a  & \phn98 & 5.1 & \phn69 & 3 & 0.8 & 2015 Oct 8\phn \\  
              &        &        &     & \phn66 & 3 & 0.8 & 2015 Oct 9\phn \\  
              &        &        &     & \phn20 & 3 & 0.8 & 2016 Sep 25    \\  
              &        &        &     & \phn30 & 1 & 0.9 & 2016 Sep 26    \\  
              &        &        &     &    120 & 4 & 0.5 & 2016 Dec 29    \\  
Andromeda~III & and3a  & \phn75 & 5.3 &    175 & 7 & 1.0 & 2015 Oct 8\phn \\  
              &        &        &     &    143 & 6 & 0.7 & 2016 Sep 7\phn \\  
Andromeda~V   & and5a  &    107 & 6.7 &    199 & 7 & 0.9 & 2015 Oct 7\phn \\  
              &        &        &     &    206 & 7 & 0.9 & 2016 Sep 27    \\  
              & and5b  &    104 & 5.0 &    203 & 8 & 0.5 & 2016 Sep 7\phn \\  
              &        &        &     & \phn97 & 3 & 0.8 & 2016 Dec 29    \\  
Andromeda~X   & and10a & \phn62 & 4.9 &    180 & 6 & 1.0 & 2015 Oct 9\phn \\  
              &        &        &     &    116 & 4 & 0.6 & 2016 Sep 28    \\  
              & and10b & \phn51 & 6.2 &    180 & 6 & 0.8 & 2016 Sep 28    \\  
              &        &        &     &    101 & 4 & 0.6 & 2016 Dec 28    \\  
              &        &        &     & \phn88 & 3 & 1.1 & 2017 Jan 1\phn \\  
\enddata
\end{deluxetable*}

We observed the five M31 dSphs with DEIMOS in the fall seasons of
2015, 2016, and 2017.  We used one slitmask on each of the three more
massive dSphs (And~VII, I, and III).  We used two slitmasks in each of
the two less massive dSphs (And~V and X) because their lower stellar
density made for sparser sampling on each slitmask.  Using two
slitmasks allowed us to increase the sample sizes.

The observations were conducted with the 1200G grating at a central
wavelength of 7800~\AA\@.  This configuration provides a spectral
range of approximately 6300--9100~\AA, but the range varies depending
on the location of the slit on the slitmask.  We used 0.7'' slits,
which yielded a spectral resolution of approximately 1.2~\AA\ FWHM\@.
The resolution was a slight function of wavelength due to changes in
anamorphic demagnification and image quality along the direction of
dispersion.  We aligned the slitmasks using at least four bright
alignment stars that were centered in 4'' boxes.  The position angles
of the slits were the same for all slits on a given slitmask, and that
angle was chosen to be close to the parallactic angle at the expected
time of observation.  This choice minimized slit losses due to
differential atmospheric refraction.  Afternoon exposures of an
internal quartz lamp were used for flat fielding, and afternoon
exposures of Ne, Ar, Kr, and Xe arc lamps (turned on simultaneously)
provided wavelength calibration.  DEIMOS's active flexure compensation
system minimized shifts of the image on the detector due to the
changing gravity vector during the observations.

Table~\ref{tab:obs} gives the exposure times for each slitmask,
separated by date of observation.  The table also gives the total
number of slits for each slitmask.  That number excludes alignment
stars, but it includes dSph non-members and slits that yielded no
useful spectra.

The raw spectra were reduced with the spec2d pipeline
\citep{coo12,new13}.  The pipeline traces the edges of each slit using
the quartz lamp exposures.  After excising the two-dimensional,
spectrally dispersed image of the slit in each of the calibration and
on-sky frames, the pipeline computed a flat field correction from the
quartz lamp exposure.  The pipeline traced the arc lines across the
slit to provide a two-dimensional wavelength solution for the slit.
The flat field was applied to the on-sky images.  Images obtained
within the same observing run were coadded into a single image of the
slit.  Then, the spectrum of the object was extracted in a boxcar
window.

The heliocentric correction can change by up to 0.4~km~s$^{-1}$ per
day in the direction of M31.  Furthermore, the reference frame for
flexure compensation in DEIMOS drifts from night to night.  It is
difficult to maintain the same reference frame for more than a week.
For these reasons, we limited two-dimensional coaddition to images
taken within the same observing run.  The observing runs spanned no
more than 3 nights in September, when the heliocentric correction
changes the most from night to night, and 5 nights in
December/January, when the heliocentric correction changes by only
0.1~km~s$^{-1}$ per night.  We extracted one-dimensional spectra from
each of these subsets.  Then, we coadded the one-dimensional spectra
in the heliocentric frame using inverse variance weighting.  The
resulting spectrum is called $s_{\rm helio}$.  We also made a separate
version of the coaddition in the geocentric frame.  The resulting
spectrum ($s_{\rm geo}$) was used to determine the slit centering
correction (Section~\ref{sec:velmsmts}).

Some stars in And~V and And~X were observed on two separate slitmasks.
Their one-dimensional spectra were coadded in the same manner
described in the previous paragraph.  The total exposure times for
these stars were 11.7 hours (And~V) and 11.1 hours (And~X)\@.

We estimated the signal-to-noise ratio (SNR) for each spectrum with
the following procedure.  We calculated the absolute deviation from 1
of the continuum-normalized spectra in the ``continuum regions''
defined by \citet{kir08a}.  These regions are defined from a the model
spectrum of a star with $T_{\rm eff} = 4300$~K, $\log g = 1.5$, and
${\rm [Fe/H]} = -1.5$.  The continuum regions are contiguous spectral
windows with widths of at least 0.5~\AA\@.  In order to qualify as a
continuum region, every pixel in the window must have a flux decrement
less than 4\% when smoothed to the spectral resolution of DEIMOS's
1200G diffraction grating.  Within the contiuum regions, we discarded
pixels that exceeded three times the median deviation in order to
discount pixels still affected by absorption lines in the star.  The
SNR was defined as the inverse of the median deviation of the
remaining pixels.  The SNR of our sample ranges from \minsn~\AA$^{-1}$
to \maxsn~\AA$^{-1}$ with a median of \medsn~\AA$^{-1}$.  To convert
to SNR per pixel, the SNR per \AA\ can be multiplied by $\sqrt{0.33}$,
where 0.33 is the number of \AA\ per pixel.

\begin{figure}
\centering
\includegraphics[width=\linewidth]{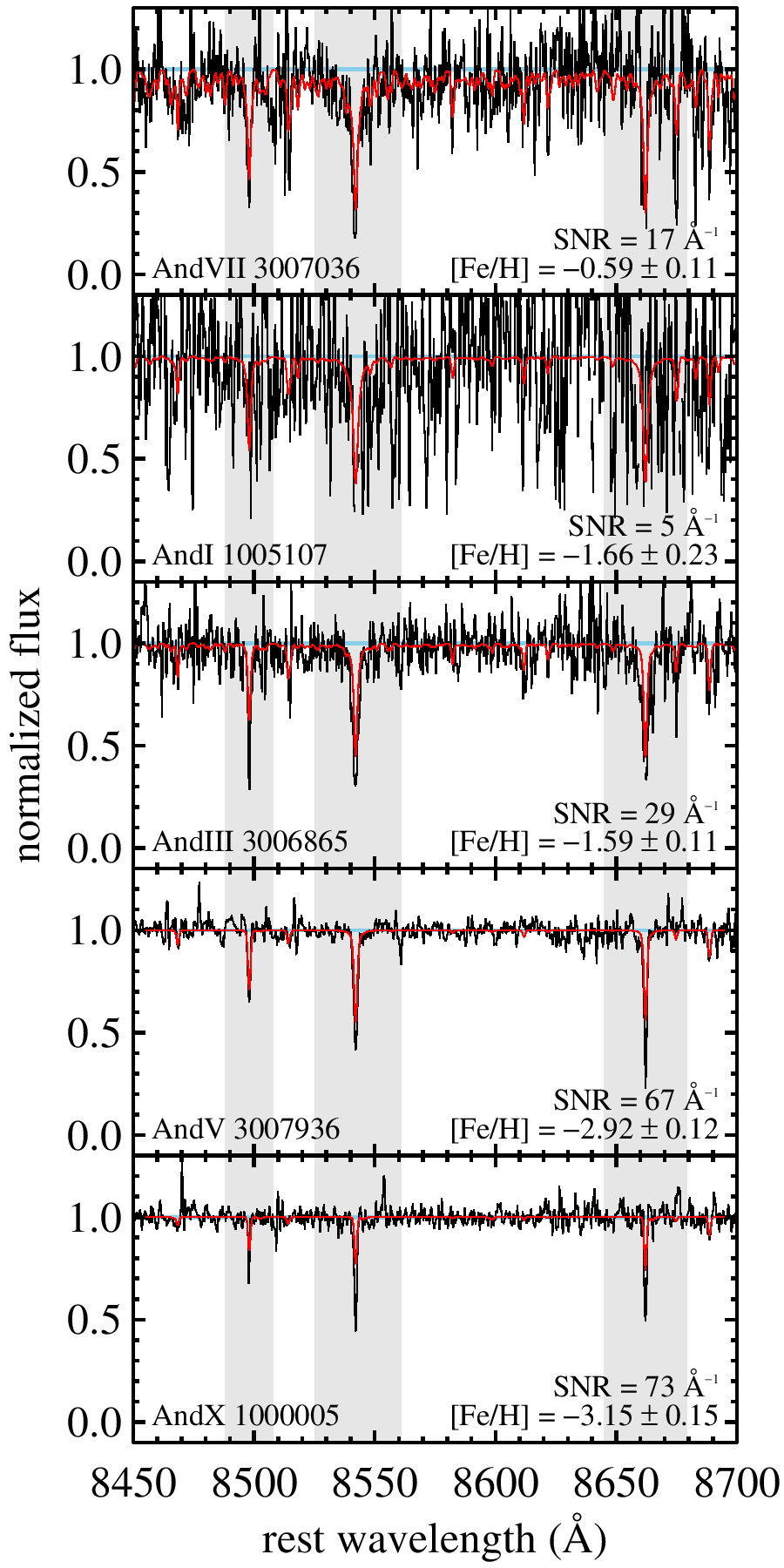}
\caption{Portions of example DEIMOS spectra in each M31 dSph.  The
  spectra shown are the star with the highest [Fe/H] in And~VII, the
  spectrum in And~I with the lowest SNR that still permitted a
  measurement of [Fe/H], the spectrum in And~III with the median SNR,
  the spectrum in And~V with the highest SNR, and the spectrum in
  And~X with the lowest [Fe/H]\@.  The red curve shows the
  best-fitting synthetic spectrum.  The \ion{Ca}{2} triplet (gray
  shading) is not well modeled and excluded from the abundance
  determination.  The full wavelength range is approximately
  6300--9100~\AA\@.\label{fig:spectra}}
\end{figure}

Figure~\ref{fig:spectra} shows some example spectra at the extremes of
metallicity and SNR\@.  The spectra are shown in only a small region
of the full wavelength range.  Although the figure shows the
best-fitting synthetic spectra, the majority of the abundance
information falls outside of this spectral range.  In particular, the
\ion{Ca}{2} triplet shown in Figure~\ref{fig:spectra} is not used to
measure abundances, but it is used to measure radial velocities.

\subsection{Velocity Measurements}
\label{sec:velmsmts}

We measured radial velocities using the procedures previously
described by \citet{kir15b}, who based their method on that of
\citet{sim07}.  We summarize the method here.

We measured velocities by cross-correlating the one-dimensional
spectrum ($s_{\rm helio}$) with template spectra observed with DEIMOS
by \citet{kir15b}.  Nine radial velocity standard stars (mostly
metal-poor giants) comprise the template library.  The velocities are
determined by minimizing $\chi^2$ between the target and template
spectra while the redshift of the target spectrum was varied.  The
entire spectrum except for regions of strong telluric absorption was
used in the cross correlation.  All of the velocity measurements were
confirmed by visual inspection.  In some cases, the automated cross
correlation failed, often because incompletely subtracted sky emission
lines affected the measurement.  In these cases, the velocity was
re-measured with a wavelength range restricted approximately to the
\ion{Ca}{2} triplet (8450--8700~\AA\ in the observer frame).

The velocity measurements can be affected by the position of the star
in the slit.  If the center of the star is misaligned perpendicular to
the slit (along the dispersion axis), then the stellar spectrum will
have a different wavelength zeropoint than the arc line spectrum.  The
reason is that the arc lines completely fill the slit, whereas the
stellar spectrum has a peaked profile that does not fill the slit
uniformly.  We corrected the wavelength zeropoint by calculating the
apparent velocity shift of telluric absorption features in the
spectrum \citep[see][]{soh07}.  Because the telluric features are
absorbed from the starlight, they have the same zeropoint as the
starlight.  We measured the ``velocity'' of the telluric features
using the same cross correlation method described in the previous
paragraph.  We used the spectrum coadded in the geocentric frame
($s_{\rm geo}$) rather than the heliocentric frame because the
telluric features should be at rest in the geocentric frame, and the
transformation from geocentric to heliocentric frames varies across
the observations.  We cross correlated this spectrum with the template
spectrum of a hot star, also observed by \citet{kir15b}.  The
wavelength range of the cross correlation was restricted to the
regions of strong telluric absorption (6866--6912~\AA, 7167--7320~\AA,
7593--7690~\AA, and 8110--8320~\AA), which include the Fraunhofer A
and B bands.  The telluric correction was subtracted from the stellar
radial velocity determined above to determine the final heliocentric
velocity.

Our determinations of the slit centering correction were based on
spectra coadded from multiple nights.  In principle, the position of
the star in the slit could change from night to night.  Therefore, it
would be best if we could measure the slit centering correction for
each night of observation.  However, the SNR from a single night is
generally too low for a reliable measurement.  Fortunately, the
majority of the mis-centering results from astrometric errors.  We
used alignment stars with coordinates in the same astrometric system
as the target stars.  However, relative astrometric errors between the
alignment and target stars will lead to slit mis-centering that is
fairly constant from night to night.  Proper motions could also cause
the bright (presumably nearby, compared to M31) alignment stars to
drift from the time their positions were measured.  These errors are
not necessarily constant in time, but they are confined to the small
angular distance the stars moved in $\sim 1$~year, which is the
approximate length of time between the first and last observations of
any particular slitmask.

We estimated errors on the velocity through Monte Carlo sampling.  We
resampled the flux of each pixel from a Gaussian with a width
corresponding to the square root of the variance estimated from the
data reduction pipeline.  We then recomputed the heliocentric velocity
and its telluric correction.  The random velocity error was taken to
be the standard deviation of $10^3$ samples.

The error also has a systematic component.  \citet{kir15b} determined
that the systematic error is 1.49~km~s$^{-1}$ from repeat measurements
of the same stars.  The final velocity error is equal to the random
error added in quadrature with the systematic error.

\subsection{Abundance Measurements}
\label{sec:abundmsmts}

We measured elemental abundances using spectral synthesis in local
thermodynamic equilibrium (LTE) in the same manner as
\citet{kir08a,kir09,kir10}.  This section gives a summary of the
procedure.

We prepared the observed spectra for abundance measurement by telluric
correction, shifting to the rest frame, and continuum normalization.
We constructed a telluric absorption template from the
continuum-normalized spectrum of a hot star.  This spectrum was raised
to the ratio of the airmasses of the target star and the hot star in
order to account for different amounts of atmospheric absorption
between the telluric and target spectra.  Then, the target star's
spectrum was divided by this scaled telluric spectrum.  Next, the
spectrum was shifted into the rest frame according to the velocity
determined in Section~\ref{sec:velmsmts}.  Finally, we made an initial
attempt at determining the continuum.  We fit a spline to the
``continuum regions'' defined by \citet{kir08a}.  The spline had a
breakpoint spacing of 100 pixels.  The spectrum was divided by the
spline.

We then searched for the best-fitting synthetic spectrum among a grid
of synthetic spectra \citep{kir11d}.  The spectra were computed with
MOOG \citep{sne73,sne12} coupled with ATLAS9 \citep{kur93} model
atmospheres and \citeposs{kir08a} line list.  We searched the grid in
effective temperature ($T_{\rm eff}$) and metallicity ([Fe/H]) using
the MPFIT implementation \citep{mar12} of Levenberg--Marquardt
$\chi^2$ minimization.  The surface gravity ($\log g$) was fixed at
the photometric isochrone value assuming distance moduli of $m-M =
24.51$ for And~I, $24.38$ for And~III \citep{ski17}, $24.44$ for
And~V, $24.41$ for And~VII \citep{mcc05}, and $24.26$ for And~X
\citep{wei19}.  The spectral range for calculating $\chi^2$ was
restricted to regions immediately surrounding iron and $\alpha$
element absorption lines.  After $T_{\rm eff}$ and [Fe/H] were
determined, we searched the grid again, this time restricting the
spectral range to $\alpha$ element absorption lines and varying only
[$\alpha$/Fe].  The value of [$\alpha$/Fe] was used for both the
ATLAS9 model atmosphere and the MOOG synthesis.

We then refined the determination of the continuum.  A synthetic
spectrum was constructed from the initial measurements of $T_{\rm
  eff}$, $\log g$, [Fe/H], and [$\alpha$/Fe].  The observed spectrum
was divided by this synthetic spectrum, and a spline with a breakpoint
spacing of 150 pixels was fit to the quotient.  The spline was fit to
the entire spectrum excluding regions of strong telluric absorption
and some strong absorption lines that were not well modeled,
including H$\alpha$ and the \ion{Ca}{2} triplet.
\footnote{The core of each line in the \ion{Ca}{2} triplet forms high
  in the photosphere or chromosphere, where the LTE assumption may not
  be valid.  For most stars, the majority of the triplet's equivalent
  width comes from the damping wings.  Therefore the abundance depends
  strongly on how well the damping constants are known.  For these
  reasons, abundances measured with the triplet are usually based on
  scaling relations, which are either empirical
  \citep[e.g.,][]{rut97,bat08a} or based on non-LTE modeling
  \citep{sta10}.  We exclude the triplet because our synthetic spectra
  assume LTE\@.} We divided the observed spectrum by this spline and
repeated the measurements of $T_{\rm eff}$, [Fe/H], and [$\alpha$/Fe].
This procedure was iterated until $T_{\rm eff}$ changed by less than
1~K and [Fe/H] and [$\alpha$/Fe] each changed by less than 0.001~dex.

We then measured individual elemental abundances.  We made a final
measurement of [Fe/H] by fixing $T_{\rm eff}$ and [$\alpha$/Fe] and
then performing a grid search restricted only to Fe absorption lines.
Abundances of Mg, Si, Ca, and Ti were measured in a similar manner.
We used the same model atmosphere for all of the individual abundance
measurements.  That means that the value of [$\alpha$/Fe] used to pick
the ATLAS9 model atmosphere was fixed at the value determined after
the last step of the continuum iteration.  The individual abundances
were determined by varying their values within MOOG\@.

When we refer to [$\alpha$/Fe], we mean the value of [$\alpha$/Fe]
used in the model atmosphere to determine all of the individual
abundances.  This value is informed by a combination of Mg, Si, Ca,
and Ti lines.  We separately refer to the individual element ratios as
[Fe/H], [Mg/Fe], [Si/Fe], [Ca/Fe], and [Ti/Fe].

\citetalias{var14a} measured [Fe/H] and [$\alpha$/Fe] for many of the
stars in our sample.  They used shallower DEIMOS spectroscopy.  The
advantage of our sample is the higher SNRs of individual spectra,
which result in more precise abundance measurements and the ability to
measure individual element ratios.  Appendix~\ref{sec:vargas} contains
a comparison of our measurements with those of \citetalias{var14a}\@.

\subsection{Membership}
\label{sec:membership}

We could not always fill the slitmask with candidate member stars.
Therefore, we designed the slitmasks to include more targets than we
expected to be members of the dSphs.  In some cases, simply to fill
available space for slits, we included targets that we knew in advance
to be non-members on the basis of their positions in the CMD\@.  In
addition to CMD position, we also considered the strength of the
\ion{Na}{1} doublet at 8190~\AA\ and radial velocity as membership
criteria.

The spectra revealed some objects to be galaxies rather than
individual stars.  The majority of these galaxies were identified by
the presence of redshifted emission lines, such as [\ion{O}{2}] at
3727~\AA\ in the rest frame.  There were also two absorption-line
galaxies in the direction of And~X whose spectra showed redshifted
Ca~H and K in absorption.

We selected dSph members on the basis of their CMD positions by
drawing a wide region around the corresponding RGB tailored for each
dSph (gray polygons in Figure~\ref{fig:cmd}).  The boundaries of the
region were subjective, but the region was chosen to be as inclusive
as reasonable to avoid a color (i.e., metallicity) bias in the target
selection.  The large majority of stars rejected on the basis of CMD
position were also rejected for at least one other reason.  A total of
\ncmdonly\ stars across all dSphs were rejected on the basis of their
CMD position alone.  These stars are bluer, redder, or brighter than
the RGB in the relevant dSph.

The \ion{Na}{1} doublet at 8190~\AA\ is sensitive to surface gravity.
It is strong in dwarf stars and weak in giants.  As a result, the
presence of a strong doublet would indicate that the star is a
foreground dwarf rather than a giant at the distance of M31.
\citet{kir12a} found that a summed equivalent width exceeding
1~\AA\ indicated a surface gravity in excess of $\log g = 4.5$.  Such
an equivalent width is easily visible by eye in our spectra, and it is
accompanied by strong damping wings.  We inspected all spectra
visually.  Stars displaying a strong \ion{Na}{1} doublet were flagged
as non-members.  The \ion{Mg}{1} line at 8807~\AA\ is also a good
surface gravity diagnostic \citep{bat12}, but we do not use it because
some of our spectra do not reach that wavelength, and the line is
usually weaker than the \ion{Na}{1} doublet.

We also used radial velocity as a membership criterion.
Section~\ref{sec:sigmav} describes our procedure for determining the
mean velocity and velocity dispersion ($\sigma_v$) of each dSph.  Any
star more than $3\sigma_v$ from the mean velocity was considered a
non-member.

\begin{splitdeluxetable*}{llcccccccBcccccccccl}
\tablewidth{0pt}
\tablecolumns{19}
\tablecaption{Catalog of Abundances\label{tab:catalog}}
\tablehead{\colhead{DSph} & \colhead{Star ID} & \colhead{RA} & \colhead{Dec} & \colhead{$V_0$} & \colhead{$I_0$} & \colhead{Slitmask} & \colhead{S/N} & \colhead{$v_{\rm helio}$} & \colhead{Member?} & \colhead{$T_{\rm eff}$} & \colhead{$\log g$} & \colhead{[Fe/H]} & \colhead{[$\alpha$/Fe]} & \colhead{[Mg/Fe]} & \colhead{[Si/Fe]} & \colhead{[Ca/Fe]} & \colhead{[Ti/Fe]} & \colhead{Comment} \\
\colhead{} & \colhead{} & \colhead{(J2000)} & \colhead{(J2000)} & \colhead{(mag)} & \colhead{(mag)} & \colhead{} & \colhead{(\AA$^{-1}$)} & \colhead{(km~s$^{-1}$)} & \colhead{} & \colhead{(K)} & \colhead{(cm~s$^{-2}$)} & \colhead{} & \colhead{} & \colhead{} & \colhead{} & \colhead{} & \colhead{} & \colhead{}}
\startdata
And~VII & 7000796 & 23:25:44.96 & $+$50:40:17.7 & $23.06 \pm 0.06$ & $21.75 \pm 0.04$ & and7a   & \phn\phn 5.2 & $-343.31 \pm  4.81$\phn & Y & $4258 \pm  51$ &    1.10 & $-1.62 \pm 0.30$ &          \nodata &          \nodata &          \nodata &          \nodata &          \nodata &                                                   \\
And~VII & 7000745 & 23:25:45.78 & $+$50:41:22.9 & $22.43 \pm 0.04$ & $21.13 \pm 0.03$ & and7a   &     \phn10.7 & $-313.66 \pm  3.52$\phn & Y & $4277 \pm  50$ &    0.86 & $-2.17 \pm 0.23$ &          \nodata &          \nodata &          \nodata &          \nodata &          \nodata &                                                   \\
And~VII & 7000663 & 23:25:46.72 & $+$50:40:22.3 & $23.10 \pm 0.07$ & $21.75 \pm 0.04$ & and7a   & \phn\phn 5.0 &     $-303.05 \pm 12.77$ & Y & $4187 \pm  54$ &    1.07 & $-1.46 \pm 0.25$ &          \nodata &          \nodata &          \nodata &          \nodata &          \nodata &                                                   \\
And~VII & 7000436 & 23:25:49.01 & $+$50:40:30.4 & $22.86 \pm 0.05$ & $21.73 \pm 0.03$ & and7a   & \phn\phn 7.0 & $-309.94 \pm  7.08$\phn & Y &        \nodata & \nodata &          \nodata &          \nodata &          \nodata &          \nodata &          \nodata &          \nodata &                                                   \\
And~VII & 7000338 & 23:25:49.62 & $+$50:41:26.2 & $22.56 \pm 0.04$ & $21.19 \pm 0.02$ & and7a   & \phn\phn 9.7 & $-309.78 \pm  5.11$\phn & Y & $4181 \pm  40$ &    0.83 & $-2.35 \pm 0.29$ &          \nodata &          \nodata &          \nodata &          \nodata &          \nodata &                                                   \\
And~VII & 7000349 & 23:25:50.32 & $+$50:40:15.5 & $22.30 \pm 0.03$ & $21.07 \pm 0.02$ & and7a   &     \phn12.1 & $-297.95 \pm  3.73$\phn & Y & $4413 \pm  39$ &    0.90 & $-2.13 \pm 0.23$ & $+0.64 \pm 0.37$ &          \nodata &          \nodata &          \nodata &          \nodata &                                                   \\
And~VII & 3013816 & 23:25:54.65 & $+$50:38:34.9 & $22.63 \pm 0.05$ & $21.20 \pm 0.02$ & and7a   &     \phn13.8 & $-321.66 \pm  2.92$\phn & Y & $4125 \pm  32$ &    0.80 & $-1.43 \pm 0.13$ &          \nodata &          \nodata &          \nodata &          \nodata &          \nodata &                                                   \\
And~VII & 3013989 & 23:26:02.55 & $+$50:40:30.0 & $22.23 \pm 0.04$ & $20.68 \pm 0.02$ & and7a   &     \phn18.3 & $-328.35 \pm  3.12$\phn & Y & $3991 \pm  22$ &    0.51 & $-1.84 \pm 0.13$ & $+0.22 \pm 0.25$ &          \nodata &          \nodata &          \nodata & $+0.08 \pm 0.21$ &                                                   \\
And~VII & 3012195 & 23:26:03.95 & $+$50:38:16.1 & $22.21 \pm 0.03$ & $20.77 \pm 0.02$ & and7a   &     \phn22.5 & $-321.78 \pm  2.55$\phn & Y & $4113 \pm  22$ &    0.62 & $-1.85 \pm 0.13$ & $+0.72 \pm 0.18$ &          \nodata &          \nodata &          \nodata & $+0.70 \pm 0.18$ &                                                   \\
And~VII & 3013609 & 23:26:04.89 & $+$50:39:32.7 & $22.10 \pm 0.03$ & $20.72 \pm 0.02$ & and7a   &     \phn25.4 & $-297.24 \pm  2.10$\phn & Y & $4201 \pm  23$ &    0.64 & $-1.59 \pm 0.11$ & $-0.04 \pm 0.21$ &          \nodata &          \nodata &          \nodata & $-0.05 \pm 0.20$ &                                                   \\
\enddata
\tablecomments{(This table is available in its entirety in machine-readable form.)}
\end{splitdeluxetable*}

Table~\ref{tab:catalog} presents the coordinates, extinction-corrected
magnitudes, SNR, velocity, elemental abundances, and membership
information for each observed star.  Stars observed on two slitmasks
are identified with the letters ``ab.''  For example, ``and5ab'' means
that the star was observed on the slitmasks and5a and and5b.  Only
abundances with errors less than 0.4~dex are given.  The last column
of the table gives comments for some stars as well as reasons for
excluding a star from membership consideration.  The codes ``CMD,''
``v,'' and ``Na'' mean that a star was ruled a non-member because of
its CMD position, velocity, or \ion{Na}{1} doublet strength,
respectively.  Many stars have multiple codes because they displayed
multiple non-membership traits.  The code ``G'' means that the object
is a galaxy, not a star.  Stars are also marked for displaying CN or
${\rm C}_2$ absorption (``C'') or TiO absorption (``TiO''), although
these were not considered in the determination of membership.
Although our synthetic spectra do have CN and ${\rm C}_2$ features,
they lack TiO\@.  For this reason, \citet{gil19} and
\citet{esc19a,esc19b} treated TiO stars as a special case.  However,
only one star in our sample (And~VII 3009310) has both TiO absorption
and a measurement of [Fe/H]\@.  Therefore, we do not consider TiO a
significant source of bias in our sample.


\section{Kinematics}
\label{sec:v}

In this section, we quantify the dynamical masses of five M31 dSphs
and examine their constituent members for evidence of rotation.
Unfortunately, the sample sizes are insufficient to test for kinematic
substructure, such as multiple chemodynamical populations
\citep[e.g.][]{wal11}.  Sample sizes of at least several hundreds are
required for this purpose \citep[e.g.,][]{bat08b,amo12,pac14}.

\subsection{Velocity Dispersion and Mass}
\label{sec:sigmav}

\begin{figure*}
\centering \includegraphics[width=\linewidth]{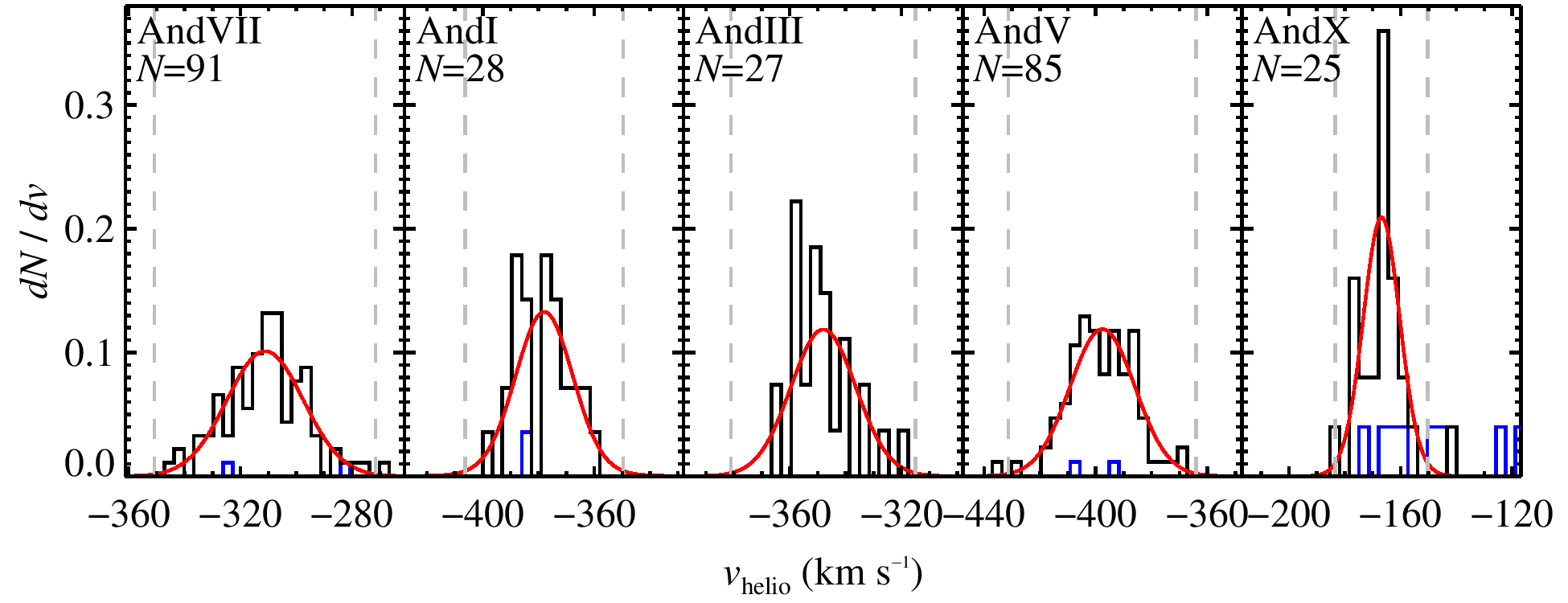}
\caption{Histograms of heliocentric radial velocities of individual
  stars in each dSph.  Stars that fail any membership criterion other
  than radial velocity are represented by the blue histograms.  The
  black histograms show the remaining potential members.  The
  histograms are normalized to enclose the same area in each panel
  regardless of the number of measurements.  The red curves show
  Gaussians with the measured $\sigma_v$ for each galaxy.  The curves
  are convolved with a function to approximate the widening induced by
  observational uncertainty.  This function is the sum of $N$
  equal-area Gaussians, where $N$ is the number of stars in the
  histogram.  The width, $\sigma_i$, of each Gaussian is the
  measurement uncertainty of star $i$.  The dashed gray vertical lines
  show the bounds of the velocity membership criterion.  Each panel
  gives the number of members ($N$) in the dSph.
  Section~\ref{sec:velmsmts} describes the radial velocity
  measurements, and Section~\ref{sec:sigmav} describes how we quantify
  the resulting velocity distributions.\label{fig:vhist}}
\end{figure*}

\begin{deluxetable*}{lccccccccc}
\tablewidth{0pt}
\tablecolumns{10}
\tablecaption{Kinematical Properties of M31 dSphs\label{tab:v}}
\tablehead{\colhead{DSph} & \colhead{$M_*/L_V$\tablenotemark{a}} & \colhead{$\log M_*$\tablenotemark{b}} & \colhead{$N_{\rm mem}$} & \colhead{$\langle v_{\rm helio} \rangle$} & \colhead{$\sigma_v$} & \colhead{$r_{1/2}$\tablenotemark{c}} & \colhead{$\log M_{1/2}$} & \colhead{$(M/L)_{1/2}$\tablenotemark{d}} & \colhead{$|v_{\rm rot}|$\tablenotemark{e}} \\
\colhead{} & \colhead{($M_{\sun}/L_{\sun}$)} & \colhead{($M_{\sun}$)} & \colhead{} & \colhead{(km~s$^{-1}$)} & \colhead{(km~s$^{-1}$)} & \colhead{(pc)} & \colhead{($M_{\sun}$)} & \colhead{($M_{\sun}/L_{\sun}$)} & \colhead{(km~s$^{-1}$)}}
\startdata
And~VII &                  0.9 &                  $7.21 \pm 0.12$ &  91 &       $-311.2 \pm 1.7$ & $13.2_{-1.1}^{+1.2}$ & $980 \pm  50$ &             $8.08 \pm 0.08$ &       $ 13 \pm   4$ & \phn$ <  7.9$ \\
And~I   &                  1.6 &                  $6.86 \pm 0.40$ &  28 & $-378.0_{-2.3}^{+2.4}$ & $ 9.4_{-1.5}^{+1.7}$ & $830 \pm  40$ &      $7.71_{-0.14}^{+0.16}$ &       $ 23 \pm  22$ & \phn$ <  9.0$ \\
And~III &                  1.8 &                  $6.27 \pm 0.12$ &  27 & $-348.0_{-2.6}^{+2.7}$ & $11.0_{-1.6}^{+1.9}$ &         $520$ &      $7.65_{-0.13}^{+0.15}$ & $ 86_{- 34}^{+ 38}$ & \phn$ <  9.5$ \\
And~V   &                  1.1 &                  $5.81 \pm 0.12$ &  85 &       $-397.7 \pm 1.5$ & $11.2_{-1.0}^{+1.1}$ & $450 \pm  20$ &             $7.59 \pm 0.08$ & $132_{- 43}^{+ 44}$ & \phn$ <  6.1$ \\
And~X   & 1.6\tablenotemark{f} &                  $5.08 \pm 0.03$ &  25 &       $-166.9 \pm 1.6$ & $ 5.5_{-1.2}^{+1.4}$ & $310 \pm  30$ &      $6.82_{-0.20}^{+0.22}$ & $177_{- 81}^{+ 92}$ & \phn$ <  5.5$ \\
\enddata
\tablenotetext{a}{Stellar mass-to-light ratio measured by \citet{woo08}.}
\tablenotetext{b}{Stellar mass, calculated from the product of luminosity measured by \citet{tol12} and stellar mass-to-light ratio from the previous column.}
\tablenotetext{c}{3-D de-projected half-light radius, taken from \citet{tol12}.}
\tablenotetext{d}{Dynamical mass-to-light ratio within the half-light radius, where the luminosity is taken from \citet{tol12}.}
\tablenotetext{e}{98\% confidence level upper limit.}
\tablenotetext{f}{\citet{woo08} did not calculate $M_*/L$ for And~X\@.  We assume the average value for dSphs measured by \citet{woo08}.}
\end{deluxetable*}

Figure~\ref{fig:vhist} shows the velocity distributions for each
galaxy.  Each dSph shows an obvious peak in radial velocity.  It is
also apparent that the velocity dispersion ($\sigma_v$) is the highest
for the dSph with the highest stellar mass (And~VII) and lowest for
the dSph with the lowest stellar mass (And~X)\@.

We measured the mean velocity ($\langle v_{\rm helio} \rangle$) and
$\sigma_v$ for each dSph using maximum likelihood.  We closely
followed the procedure of \citet{wal06}.  The likelihood function is

\begin{equation}
L = \prod_i \frac{1}{\sqrt{2\pi(\sigma_v^2 + \delta v_{{\rm helio},i}^2)}} \exp \frac{-(v_{{\rm helio},i} - \langle v_{\rm helio} \rangle)^2}{2(\sigma_v^2 + \delta v_{{\rm helio},i}^2)} \label{eq:l}
\end{equation}

\noindent The product operates over all member stars $i$.  The
heliocentric velocity of star $i$ is $v_{{\rm helio},i}$, and its
uncertainty is $\delta v_{{\rm helio},i}$.  The free parameters are
$\langle v_{\rm helio} \rangle$ and $\sigma_v$.

We found the maximum likelihood with a Monte Carlo Markov chain (MCMC)
using a Metropolis algorithm.  We adopted the initial values of
$\langle v_{\rm helio} \rangle$ and $\sigma_v$ from \citet{tol12}, who
measured the kinematics of M31 satellites from shallower DEIMOS
spectra.  The proposal density of the Metropolis algorithm was a
normal distribution with $\sigma = 2.5$~km~s$^{-1}$.  The chain had
$10^6$ links plus an initial $10^5$ links, which were discarded as
burn-in.  Visual inspection of the chains showed that they were
well-converged after burn-in.  We took the most likely values of
$\langle v_{\rm helio} \rangle$ and $\sigma_v$ to be the medians
($50^{\rm th}$ percentile) of the distributions.  The asymmetric
$1\sigma$ errors are those values that enclose 68\% of the
distribution ($16^{\rm th}$ and $84^{\rm th}$ percentiles).

Our first attempt at measuring these values included stars whose
velocities satisfied the criterion $|v_{{\rm helio},i} - \langle
v_{\rm helio} \rangle | < 3\sigma_v$, using the $\langle v_{\rm helio}
\rangle$ measurements of \citet{tol12} as a starting guess.  After
computing new measurements of $\langle v_{\rm helio} \rangle$ and
$\sigma_v$, we re-evaluated the membership of each star using the
$3\sigma_v$ criterion.  Then, we recomputed the entire MCMC chain.  We
iterated this procedure until the list of member stars was unchanged
between iterations.

Table~\ref{tab:v} gives summary structural and kinematical information
for stars in the dSphs.  It is worth discussing the stellar mass
measurements because so many properties of the dSphs depend on stellar
mass.  We adopt the $M_*$ measurements of \citet{woo08}.  That study
estimated stellar mass in two different ways: scaling relations
between integrated galaxy colors and $M_*/L$ \citep{bel01} and
integrations of SFHs from CMDs \citep{mat98}.  Although
\citeauthor{woo08}\ did not calculate uncertainties on $M_*/L$ for
each galaxy, we can estimate the uncertainties by the scatter between
the two mass measurement methods.  The standard deviation of the
difference in mass measurements is 0.16~dex (a factor of 1.5).  The
deviations do not depend on galaxy luminosity.
\citeauthor{woo08}\ used the SFH method for the two M31 dSphs (And~I
and And~III) for which they had access to measured SFHs.  They used
the color--mass relation for And~V and And~VII\@.  They did not have
any measurements for And~X, which was discovered just prior to the
publication of their paper.  Therefore, we assume that And~X has the
average value of $M_*/L$ for the dSphs that
\citeauthor{woo08}\ included in their sample.

The values for $\langle v_{\rm helio} \rangle$ and $\sigma_v$ are
largely consistent with previous measurements
\citep{tol12,col13}.\footnote{Our sample sizes are smaller than those
  of \citet{tol12} but larger than that of \citet{col13}.  Our
  exposure times are longer than both studies by a factor of 5--10.}
The most discrepant measurement is $\langle v_{\rm helio} \rangle$ for
And~VII, for which our measurement differs by 2.3 standard deviations
from that of \citet{tol12}.  The average offset in $\langle v_{\rm
  helio} \rangle$ is 2.5~km~s$^{-1}$, where our measurements are more
negative.  The offset is probably a result of a difference in velocity
zeropoints between the radial velocity template spectra.  If we shift
all of our measurements of $\langle v_{\rm helio} \rangle$ by
2.5~km~s$^{-1}$, they agree with \citet{tol12} within 1.1 standard
deviations.  All of the measurements of $\sigma_v$ are consistent
within one standard deviation.

The dynamical mass of each dSph can be estimated using the formula of
\citet{wol10}: $M_{1/2} = 3G^{-1} \sigma_v^2 r_{1/2}$, where $r_{1/2}$
is the 3-D (de-projected) half-light radius, $M_{1/2}$ is the mass
enclosed within $r_{1/2}$, and $G$ is the gravitational constant.  We
adopt the values of $r_{1/2}$ from \citet{tol12}.  These values agree
with those of \citet{mar16c} except for And~I, where
\citeauthor{mar16c}'s measurement is larger by 31\% after correcting
for de-projection.  The value of $M_{1/2}$ would be larger by the same
factor if we used the larger half-light radius.  Our measurements of
$M_{1/2}$, shown in Table~\ref{tab:v}, are entirely consistent with
those of \citet{tol12}, which is not surprising given that
measurements of $\sigma_v$ are consistent.

These mass estimates reaffirm that the dSph satellites of M31 are
dominated by dark matter.  The dynamical mass-to-light ratios ($M/L =
2M_{1/2}/L$) are given in Table~\ref{tab:v}.  In agreement with
previous studies, the values are in excess of what would be expected
($M/L \sim 2$) for an ancient stellar population in dynamical
equilibrium with no dark matter.  The ratios of total mass to stellar
mass within the half-light radii ($2M_{1/2}/M_*$) range from
$\minmratio_{-\minmratioerrlo}^{+\minmratioerrhi}$ (\minmratiodsph) to
$\maxmratio_{-\maxmratioerrlo}^{+\maxmratioerrhi}$ (\maxmratiodsph)\@.
The ratio for a dSph devoid of dark matter would be $\sim 1$\@.

Stars in binary systems exhibit radial velocity variability, which
inflates measurements of $\sigma_v$.  The effect is to spuriously
increase mass estimates because the increase in $\sigma_v$ does not
reflect the depth of the galaxy's gravitational potential.  The best
practice is to identify stars with variable velocities via multiple
epochs of spectroscopy.  In principle, we could have used previous
SPLASH observations and subsets of our own observations to quantify
binarity.  However, binarity negligibly affects $\sigma_v$ for
galaxies with $\sigma_v \sim 10$~km~s$^{-1}$ \citep{spe17}.
\citet{min10} also showed that binarity does not significantly affect
mass estimates of classical dSphs.

\subsection{Rotation}
\label{sec:rot}

DSphs generally do not exhibit rotation.  \citet{whe17} found that
dSphs generally have rotation-to-dispersion ratios of $v_{\rm
  rot}/\sigma_v < 0.5$.  And~II, one of M31's dSph satellites, is a
notable exception.  It not only rotates, but it rotates at
8.6~km~s$^{-1}$ about its major axis, i.e., prolate rotation
\citep{ho12}.  The ratio of rotation to dispersion is $v_{\rm
  rot}/\sigma_v = 1.1 \pm 0.3$.  This unusual sense of rotation has
been suggested to reflect a merger of two dwarf galaxies of similar
mass \citep{amo14,lok14,fou17}.  In light of the information that
rotation possibly encodes, we tested for and quantified rotation in
the five M31 satellites in our sample.

We followed \citeposs{whe17}\ procedure for measuring rotation.  The
method is an extension of the measurement of $\sigma_v$ presented in
Section~\ref{sec:sigmav}.  Introducing rotation requires a slight
modification of the likelihood function.

\begin{eqnarray}
L &=& \prod_i \frac{1}{\sqrt{2\pi(\sigma_v^2 + \delta v_{{\rm helio},i}^2)}} \exp \frac{-(v_{{\rm helio},i} - v_i')^2}{2(\sigma_v^2 + \delta v_{{\rm helio},i}^2)} \label{eq:lrot} \\
v_i' &=& \langle v_{\rm helio} \rangle + v_{\rm rot} \cos (\theta - \theta_i) \label{eq:vprime}
\end{eqnarray}

\noindent Equation~\ref{eq:vprime} represents a flat rotation curve
with a velocity of $v_{\rm rot}$ relative to $\langle v_{\rm helio}
\rangle$.  Star $i$ has a position angle of $\theta_i$ measured with
respect to the center of the dSph \citep[coordinates taken
  from][]{tol12}.  The angle $\theta$ is the orientation of the
rotation axis.  The velocity $v_{\rm rot}$ is unconstrained, but
$\theta$ is constrained between $0^{\circ}$ and $180^{\circ}$.
\citeauthor{whe17}\ also considered an isothermal model of rotation,
but we chose to keep our test for rotation as simple as possible.  The
isothermal model might be more realistic than the flat rotation curve,
but it also benefits from a large sample out to large radius, which
our sample does not provide.

We found the posterior distributions of $\langle v_{\rm helio}
\rangle$, $\sigma_v$, $v_{\rm rot}$, and $\theta$ using the same MCMC
procedure described in Section~\ref{sec:sigmav}.  In all cases, the
inner 68\% of the posterior distribution of $v_{\rm rot}$ includes
zero.  The $|v_{\rm rot}|$ upper limits that include 95\% of all MCMC
trials for the five dSphs range from 6 to 10~km~s$^{-1}$ (see
Table~\ref{tab:v}).  Thus, we found no evidence for rotation in these
galaxies, consistent with previous studies using shallower
spectroscopy.  These non-detections reinforce that And~II is
especially unusual for exhibiting not only rotation but prolate
rotation.


\section{Metallicity}
\label{sec:feh}

In this section, we analyze the stellar metallicities of the M31
dSphs.  First, we discuss the MZR of the M31 satellites.  Then, we
interpret the MDFs through the lens of one-zone models of chemical
evolution.  We include a comparison of the M31 dSph MDFs with those of
MW dSphs.

\subsection{Metallicity Distributions}

\begin{deluxetable*}{lcccccc}
\tablewidth{0pt}
\tablecolumns{7}
\tablecaption{Chemical Properties of M31 dSphs\label{tab:abund}}
\tablehead{\colhead{DSph} & \colhead{$N_{\rm{[Fe/H]}}$} & \colhead{$\langle {\rm [Fe/H]} \rangle$} & \colhead{$\sigma({\rm [Fe/H]})$} & \colhead{median([Fe/H])} & \colhead{$N_{\rm{[\alpha/Fe]}}$} & \colhead{$\langle {\rm [\alpha/Fe]} \rangle$}}
\startdata
And~VII &  87 &        $-1.37 \pm 0.01$ & $0.36$ & $-1.39$ &  56 &        $+0.18 \pm 0.03$ \\
And~I   &  28 &        $-1.51 \pm 0.02$ & $0.34$ & $-1.47$ &  20 &        $+0.14 \pm 0.04$ \\
And~III &  24 &        $-1.75 \pm 0.03$ & $0.43$ & $-1.98$ &  21 &        $+0.11 \pm 0.04$ \\
And~V   &  81 &        $-1.84 \pm 0.01$ & $0.41$ & $-1.82$ &  57 &        $+0.07 \pm 0.03$ \\
And~X   &  21 &        $-2.27 \pm 0.03$ & $0.47$ & $-2.38$ &   9 &        $+0.27 \pm 0.09$ \\
\enddata
\end{deluxetable*}

\begin{figure}
\centering
\includegraphics[width=\linewidth]{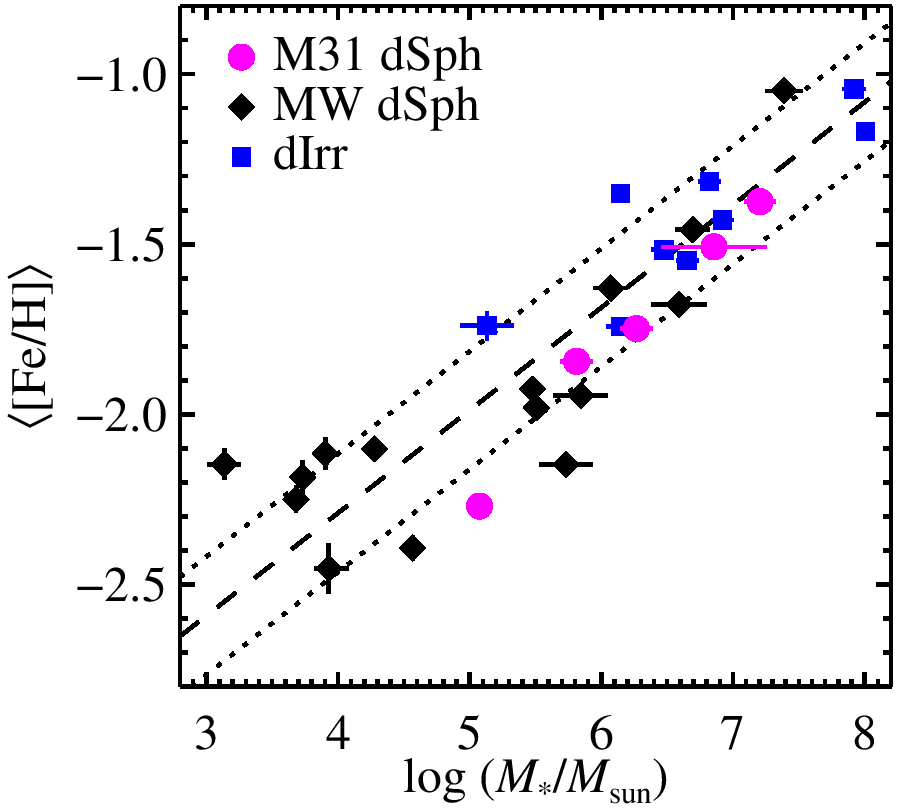}
\caption{Stellar mass--metallicity relation of Local Group dwarf
  galaxies.  The M31 dSphs are those analyzed in this paper.  The MW
  dSphs and Local Group dIrrs were analyzed by \citetalias{kir13b}.
  The dotted line shows the MZR fit to dIrrs and MW dSphs
  \citepalias{kir13b}, and the dotted lines show the $1\sigma$
  dispersion.  The M31 dSphs follow the relation.  Whereas the
  \citetalias{kir13b}'s version of this plot relied on coadded spectra
  for M31 dSphs, this plot is based entirely on metallicity
  measurements for individual stars.  Section~\ref{sec:abundmsmts}
  provides an overview of the abundance measurements, and
  Section~\ref{sec:feh} discusses the implications of the
  measurements.\label{fig:mzr}}
\end{figure}

Table~\ref{tab:abund} and Figure~\ref{fig:mzr} show the average
stellar metallicities\footnote{The quantity $\langle {\rm [Fe/H]}
  \rangle$ is the average of the logarithmic stellar metallicities,
         [Fe/H].  As pointed out by \citet{esc18}, many papers about
         galaxy simulations instead quote the logarithm of the
         average.  The distinction should be kept in mind for any
         comparisons between observed and simulated galaxy
         metallicities.} of our five M31 dSphs.  The error bars for
$\langle {\rm [Fe/H]} \rangle$ reflect the error on the mean.
Table~\ref{tab:abund} also gives the inverse variance-weighted
standard deviation and median of each MDF\@.  The mean differs from
the median both because the MDFs are asymmetric and because the mean
is weighted by inverse variance, whereas the median is not.

The metallicities increase monotonically with stellar mass, in line
with the well-established MZR\@.  Furthermore, the MZR of M31 dSphs
lies within the envelope defined by MW dSphs \citepalias{kir13b}.
And~X's metallicity lies outside of the $1\sigma$ dispersion, but it
is well in line with low-mass MW satellites.  It falls directly
between those of Ursa Minor and Hercules.  All of the M31 dSphs fall
slightly below the best-fit relation measured by \citetalias{kir13b}
but no more so than some of the MW dSphs.

\citetalias{kir13b} and \citetalias{var14a} previously measured the
average metallicities of all five dSphs in our sample.
\citetalias{kir13b} used spectral coaddition rather than measurements
of individual stars.  Despite the difference in technique, our new
measurements agree within 0.2~dex except for And~VII, which we find to
be 0.25~dex more metal-rich than \citetalias{kir13b}.
\citetalias{var14a} measured metallicities of individual stars, using
nearly the same code that we use here.  The major difference between
the two studies is that the spectra analyzed here have higher SNR\@.
Additionally, we adopt a slightly different membership selection here.
Our measurements agree with those of \citetalias{var14a} within
0.2~dex except for And~I, which we measure to be 0.40~dex more
metal-poor.  The agreement is notable because we find an offset of
$\meanfehdiff \pm \meanfehdifferr$ in [Fe/H] between our two studies
(see Appendix~\ref{sec:vargas}).  We do not correct this offset in
comparing the mean metallicities.  \citet{ho15} also presented
metallicity distributions of luminous M31 satellites, including
And~VII, using the \ion{Ca}{2} infrared triplet.  They found $\langle
{\rm [Fe/H]} \rangle = -1.30 \pm 0.07$, in close agreement with our
own measurement.

\subsection{Chemical Evolution Models}

A galaxy's gas flow history shapes its MDF\@.  Therefore, the MDF can
be used as a tool to infer its past chemical evolution.  The simplest
way to consider galactic chemical evolution is a one-zone model
\citep{pag97}.  In this model, the galaxy's gas is assumed to be
well-mixed at all times.  Therefore, it has a single composition at
any time.  The MDF shape is influenced by the yields of stellar
populations (which are usually assumed to be constant with time), the
flow of metals out of the galaxy, and the accretion of gas (usually
assumed to be metal-free).

We fit the M31 dSph MDFs with the same models used by \citet{kir11a}
and \citetalias{kir13b}.  They are a Leaky Box Model
\citep{sch63,tal71}, Pre-Enriched Model \citep{pag97}, and Accretion
Model \citep{lyn75}.  The Leaky Box Model presumes that the galaxy can
lose gas at a rate proportional to the SFR, but it does not acquire
new gas.  The only free parameter is the effective yield ($p_{\rm
  eff}$), which is the stellar yield minus metal loss due to outflow.
The Pre-Enriched Model that we use here is identical to the Leaky Box
Model except that it has an additional free parameter: the initial
metallicity (${\rm [Fe/H]}_0$).  The Accretion Model is an extension
of the Leaky Box Model where metal-free gas is allowed to fall onto
the galaxy with a specific form as a function of time.  The free
parameters are $p_{\rm eff}$ and the accretion parameter $M$, which is
the ratio of the final stellar mass to the initial gas mass.
\citetalias{kir13b} gave the equations for each model.  We fit the
models using an MCMC, as described by \citetalias{kir13b}.

\begin{figure*}
\centering
\includegraphics[width=0.945\linewidth]{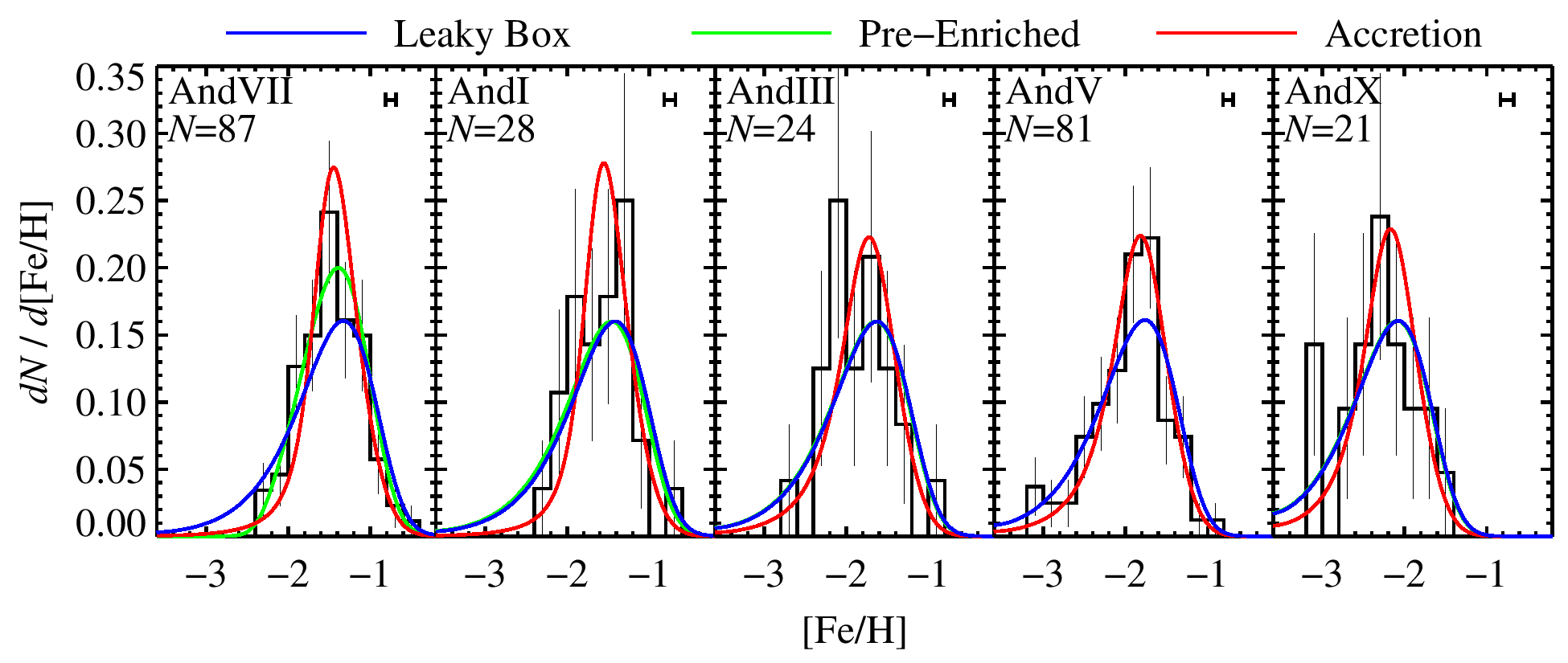}
\caption{The MDFs of M31 dSphs in bins of 0.2~dex.  Error bars are
  Poissonian.  Only stars with errors less than 0.4~dex contribute to
  the MDFs.  The error bars in the upper right of each panel show
  median uncertainties in [Fe/H]\@.  Best-fit chemical evolution
  models are shown as colored curves.  The curves are convolved with a
  function to approximate the widening induced by observational
  uncertainty, as described in the caption to Figure~\ref{fig:vhist}.
  The Pre-Enriched Model is a generalization of the Leaky Box Model,
  where the initial metallicity is allowed to be non-zero.  The
  Accretion Model is a different generalization of the Leaky Box Model
  that allows gas infall.  The green curves are not visible in the
  right three panels because the blue curves completely overlap them.
  Section~\ref{sec:feh} describes the chemical evolution models, and
  Table~\ref{tab:gce} gives the best-fit parameters of each
  model.\label{fig:fehhist}}
\end{figure*}

\begin{deluxetable*}{lcccccccccl}
\tablecolumns{11}
\tablewidth{0pt}
\tablecaption{Chemical Evolution Models\label{tab:gce}}
\tablehead{ & Leaky Box & & \multicolumn{3}{c}{Pre-Enriched} & & \multicolumn{3}{c}{Accretion} & \\ \cline{2-2} \cline{4-6} \cline{8-10}
\colhead{dSph} & \colhead{$p_{\rm eff}$\tablenotemark{a} ($Z_\sun$)} & & \colhead{$p_{\rm eff}$\tablenotemark{a} ($Z_\sun$)} & \colhead{${\rm [Fe/H]}_0$\tablenotemark{b}} & \colhead{$\Delta{\rm AICc}$\tablenotemark{c}} & & \colhead{$p_{\rm eff}$\tablenotemark{a} ($Z_\sun$)} & \colhead{$M$\tablenotemark{d}} & \colhead{$\Delta{\rm AICc}$\tablenotemark{c}} & \colhead{Best Model}}
\startdata
And~VII       & $0.049 \pm 0.006$         & & $0.038_{-0.005}^{+0.006}$ &             $-2.40_{-0.17}^{+0.13}$ &     \phs\phn $ 15.94$ & & $0.044 \pm 0.004$         &    $ 5.9_{-2.5}^{+4.2}$ &     \phn $ 10.28$ & Pre-Enriched \\
And~I         & $0.040_{-0.008}^{+0.009}$ & & $0.036_{-0.009}^{+0.010}$ &                          $<  -2.18$ &     \phn\phn $ -2.24$ & & $0.035_{-0.005}^{+0.006}$ &    $ 6.4_{-3.2}^{+4.4}$ & \phn\phn $  3.72$ & Accretion    \\
And~III       & $0.025_{-0.005}^{+0.007}$ & & $0.024_{-0.005}^{+0.007}$ &                          $<  -2.83$ &     \phn\phn $ -2.31$ & & $0.021_{-0.004}^{+0.005}$ &    $ 2.6_{-1.1}^{+2.0}$ &     \phn $ -1.83$ & Leaky Box    \\
And~V         & $0.018_{-0.002}^{+0.003}$ & & $0.018_{-0.002}^{+0.003}$ &                          $< -16.65$ &     \phn\phn $ -2.10$ & & $0.017 \pm 0.002$         &    $ 2.5_{-0.7}^{+1.1}$ & \phn\phn $  4.47$ & Accretion    \\
And~X         & $0.009_{-0.002}^{+0.003}$ & & $0.009_{-0.002}^{+0.003}$ &                          $<  -3.51$ &     \phn\phn $ -2.38$ & & $0.008 \pm 0.002$         &    $ 2.7_{-1.2}^{+2.7}$ &     \phn $ -2.21$ & Leaky Box    \\
\enddata
\tablecomments{Error bars represent 68\% confidence intervals.  Upper limits are at 95\% (2$\sigma$) confidence.}
\tablenotetext{a}{Effective yield.}
\tablenotetext{b}{Initial metallicity.}
\tablenotetext{c}{Difference in the corrected Akaike information criterion between the specified model and the Leaky Box model.  Positive numbers indicate that the specified model is preferred over the Leaky Box model.}
\tablenotetext{d}{Accretion parameter, which is the ratio of final stellar mass to initial gas mass.}
\end{deluxetable*}

Figure~\ref{fig:fehhist} shows the MDFs for each dSph along with the
best-fit chemical evolution models.  The parameters for the models are
given in Table~\ref{tab:gce}.  The dSphs are ordered left to right in
Figure~\ref{fig:fehhist} from high to low stellar mass.  Two trends
are apparent.  First, the MZR is reflected in the way that the peaks
of the MDFs shift more metal-poor with decreasing $M_*$.  Second, only
the most massive dSph, And~VII, fits the Pre-Enriched Model.  We could
measure only upper limits to ${\rm [Fe/H]}_0$ in And~I, III, V, and
X\@.  In the limit ${\rm [Fe/H]}_0 \rightarrow -\infty$, the
Pre-Enriched Model devolves into the Leaky Box Model.  For that
reason, the Pre-Enriched Model (green) is overlapped by the Leaky Box
Model (blue).

However, there is no apparent trend for which model fits best.  We
assessed the best model with the corrected Akaike information
criterion \citep[AICc][]{aka74,sug78}.  The AICc evaluates the
goodness of fit of a model with a penalty on the number of free
parameters.  More complex models must overcome this penalty to be the
favored model.  The most meaningful way to interpret the AICc is the
difference in AICc between two models: $\Delta {\rm AICc}$.
Table~\ref{tab:gce} gives $\Delta {\rm AICc}$ compared against the
Leaky Box Model.  Positive values of $\Delta {\rm AICc}$ for the
Pre-Enriched and Accretion Models indicate that they are preferred
over the Leaky Box Model.  \citeposs{jef98} scale suggests a way to
interpret $\Delta {\rm AICc}$, which is approximately twice the
natural logarithm of the ratio of the likelihoods of the two models.
Values greater than 2.3, 4.6, and 9.2 constitute ``substantial,''
``strong,'' and ``decisive'' evidence, respectively.  Between the
Pre-Enriched and Accretion Models, the preferred model has the larger
$\Delta {\rm AICc}$.  And~VII is the only model where the Pre-Enriched
Model fits best.  The Accretion Model fits And~I and And~V the best.
The Leaky Box Model is an adequate fit to And~III and And~X with no
need for any free parameters beyond $p_{\rm eff}$.

\citet{ho15} found that the Leaky Box Model is a good fit to the MDF
of And~VII, in tension with our conclusion.  Their MDF included 107
stars, which is an even larger sample than ours.  The major
differences between the two studies are the exposure times and the
measurement technique employed.  Our exposure times were at least five
times longer than those of \citet{ho15}.  We also use spectral
synthesis of \ion{Fe}{1} lines rather than the equivalent widths of
the \ion{Ca}{2} infrared triplet.  We suggest that both factors result
in smaller random and systematic errors in our measurements.  We
conclude the MDF of And~VII is inconsistent with the Leaky Box Model.

Extremely metal-poor (EMP, ${\rm [Fe/H]} < -3$) stars are interesting
because they presumably formed very early in the history of the
Universe.  Our sample contains six EMP stars: three each in And~V and
And~X\@.  The absence of EMP stars in And~VII, And~I, and And~III is
consistent with their best-fit chemical evolution models, which
predict less than one EMP star in each dSph.  The best-fit models for
And~V and And~X predict 1.6 and 1.9 EMP stars, respectively.  These
predictions are consistent within Poisson uncertainty with the three
observed EMP stars in each dSph.

\subsection{Comparison of M31 dSphs to MW dSphs}

\begin{figure*}
\centering
\includegraphics[width=0.945\linewidth]{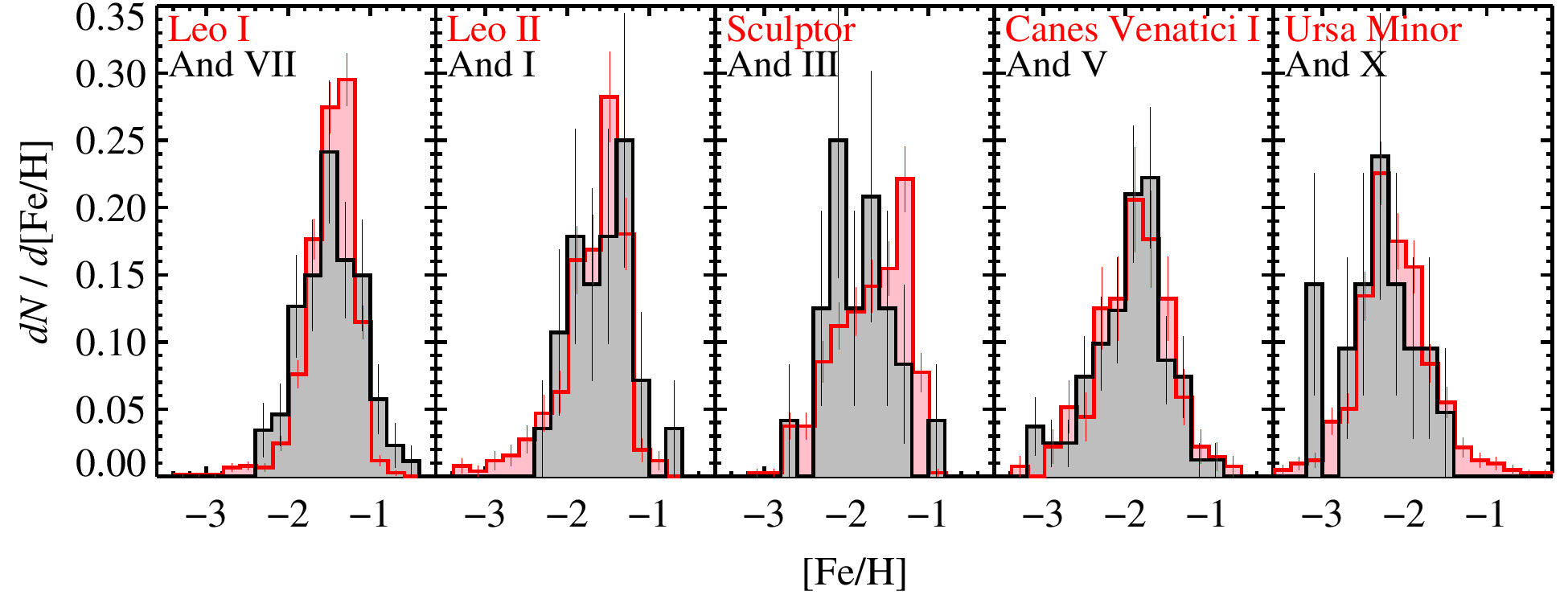}
\caption{MDFs of M31 dSphs shown with corresponding MDFs of MW dSphs
  \citep{kir10,kir11a} of similar average metallicity.  The MDFs are
  measured with identical techniques.  Error bars are Poissonian.
  Only stars with errors less than 0.4~dex contribute to the MDFs.
  Section~\ref{sec:feh} discusses the comparison between the pairs of
  MDFs.\label{fig:mwfehhist}}
\end{figure*}

It is illustrative to compare the MDFs of M31 dSphs with those of MW
dSphs of similar metallicity or stellar mass.
Figure~\ref{fig:mwfehhist} shows the M31 dSph MDFs with the MDF of the
MW dSph with the closest $\langle {\rm [Fe/H]} \rangle$ in the sample
of \citetalias{kir13b}\@.\footnote{Strictly speaking, Leo~I ($\langle
  {\rm [Fe/H]} \rangle = -1.45$) has the closest $\langle {\rm [Fe/H]}
  \rangle$ to both And~VII and And~I, but we compare And~I with the
  next closest match, Leo~II ($\langle {\rm [Fe/H]} \rangle = -1.63$),
  to show a larger diversity of MW dSph MDFs.}  We expect the MW dSph
MDFs to be slightly narrower than the M31 dSph MDFs because the
uncertainties are smaller for the MW dSphs.  Indeed, the MDF of Leo~I
is slightly narrower than that of And~VII but otherwise similar.  The
other dSph pairs show some differences.  And~I lacks very metal-poor
stars compared to Leo~II.  And~III has a symmetric MDF compared to
Sculptor, which shows a sharp metal-rich cut-off.  \citetalias{kir13b}
explained the cut-off as an effect of ram pressure stripping.  The MDF
of And~V is actually slightly narrower than that of Canes Venatici~I,
which explains why And~V favors the Accretion Model.  Gas accretion
sets up an equilibrium metallicity that causes stars to form in a
narrow range of [Fe/H]\@.  And~X lacks metal-rich stars compared to
Ursa Minor.  However, Ursa Minor is unusual for having a metal-rich
tail.

The similarity in MDF shapes extends to derived chemical evolution
model parameters.  For example, \citetalias{kir13b} found that the MDF
of Leo~I was best fit by the Accretion Model with $p_{\rm eff} = 0.043
\pm 0.001$ and $M = 7.9_{-1.0}^{+1.2}$.  Although And~VII formally
prefers the Pre-Enriched Model, its best-fit Accretion Model has
$p_{\rm eff}$ and $M$ perfectly in line with those of Leo~I\@.  The
agreement between model parameters persists for most of the other dSph
pairs, even if the preferred model is not always the same.  The
exception is And~X\@.  Its MW counterpart, Ursa Minor, has a best-fit
accretion parameter of $M = 11.0_{-4.5}^{+5.6}$, which is inconsistent
with our determination of $M$ for And~X\@.  We suggest that none of
these models is a particularly good fit to Ursa Minor because of its
metal-rich tail.

There is significant diversity in MDF shapes in MW dSphs even at
similar $\langle {\rm [Fe/H]} \rangle$.  As an example,
Figure~\ref{fig:mwfehhist} would not give the impression that the MW
and M31 dSphs agreed very well if we chose the MW dSph with the
closest $M_*$ rather than the closest $\langle {\rm [Fe/H]} \rangle$.
Given the diversity of MDF shapes and the sample sizes in the M31
dSphs, we do not have any reason to conclude that the M31 satellites
experienced a different metallicity enrichment history as a function
of stellar mass from that of the MW satellites.


\section{Alpha Element Enhancement}
\label{sec:alphafe}

In this section, we present [$\alpha$/Fe] ratios for And~I, III, V,
VII, and X\@.  These are a subset of the dSphs studied by
\citetalias{var14a}, but the spectra used here have much higher SNR\@.
As a result, we are able to measure not only a single ``alpha''
abundance but also individual element ratios, like [Si/Fe].  We
discuss qualitative trends in the [$\alpha$/Fe] vs.\ [Fe/H] diagram,
and we provide some quantitative metrics of chemical evolution.  We
also compare the distributions of abundance ratios in the M31
satellites to those of the MW satellites.

\begin{figure}
\centering
\includegraphics[width=\linewidth]{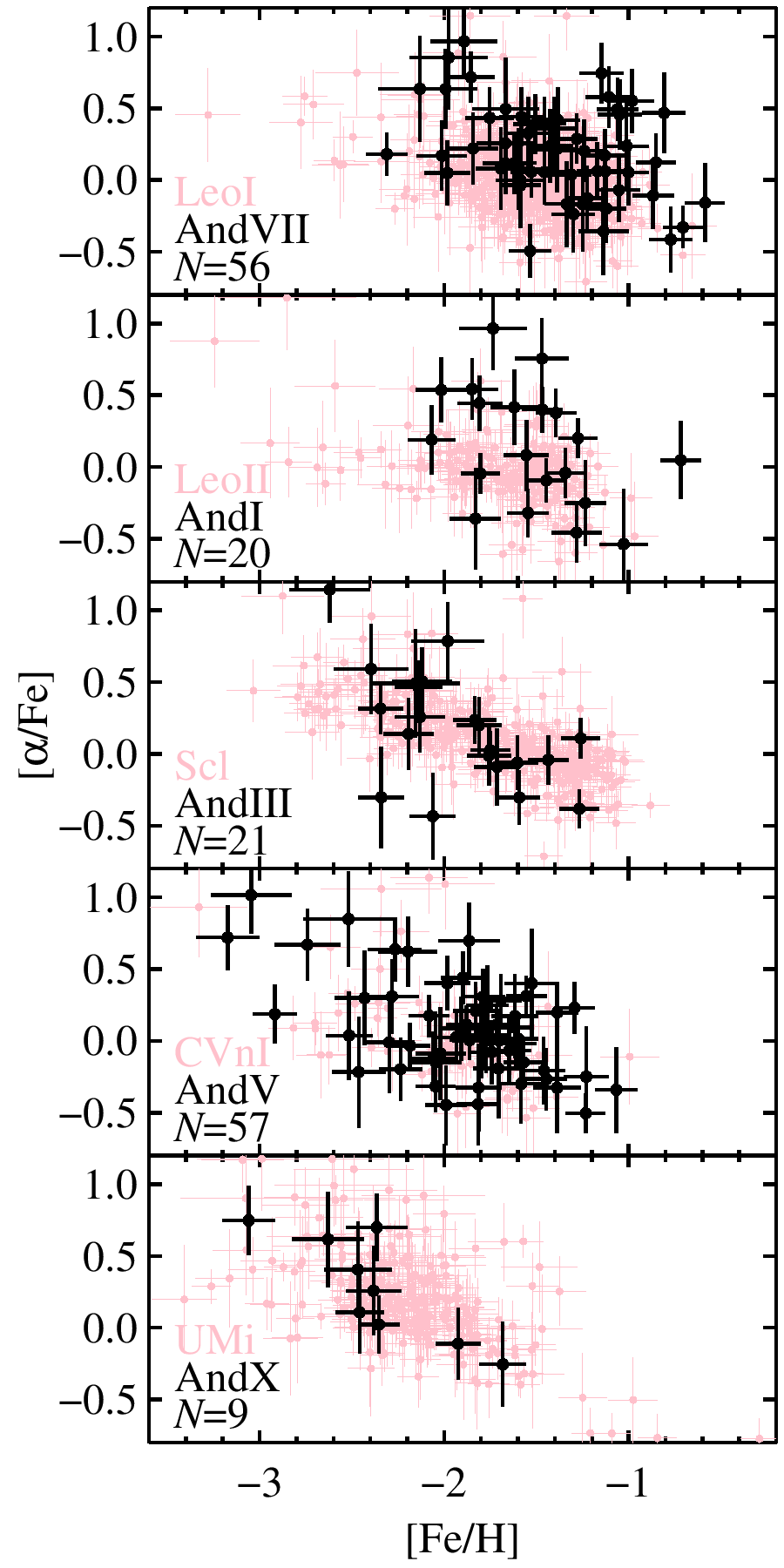}
\caption{Trends of of [$\alpha$/Fe] with [Fe/H]\@.  The M31 dSphs are
  shown in black.  In the same panel, the MW dSph with the closest
  $\langle {\rm [Fe/H]} \rangle$ is shown in pink
  \citep{kir10,kir11b}, as in Figure~\ref{fig:mwfehhist}.  The more
  massive, metal-rich dSphs have shallower slopes
  (Table~\ref{tab:alphafe}) because they maintained higher SFRs.
  Section~\ref{sec:abundmsmts} describe the abundance measurements,
  and Section~\ref{sec:alphafe} discusses their
  significance.\label{fig:alphafe}}
\end{figure}

\begin{deluxetable*}{lcccc}
\tablewidth{0pt}
\tablecolumns{5}
\tablecaption{Abundance Trends in dSphs\label{tab:alphafe}}
\tablehead{\colhead{DSph} & \colhead{$\log (M_*/M_{\sun})$\tablenotemark{a}}& \colhead{$\log (M_{1/2}/M_{\sun})$\tablenotemark{b}} & \colhead{$\langle {\rm [Fe/H]} \rangle$} & \colhead{$d{\rm [\alpha/Fe]}/d{\rm [Fe/H]}$}}
\startdata
\cutinhead{M31 dSphs}
And~VII          & $7.21 \pm 0.12$ &        $8.08 \pm 0.08$ & $-1.37 \pm 0.01$ & $-0.42_{-0.10}^{+0.09}$ \\
And~I            & $6.86 \pm 0.40$ & $7.71_{-0.14}^{+0.16}$ & $-1.51 \pm 0.02$ & $-0.96_{-0.24}^{+0.24}$ \\
And~III          & $6.27 \pm 0.12$ & $7.65_{-0.13}^{+0.15}$ & $-1.75 \pm 0.03$ & $-0.79_{-0.14}^{+0.14}$ \\
And~V            & $5.81 \pm 0.12$ &        $7.59 \pm 0.08$ & $-1.84 \pm 0.01$ & $-0.53_{-0.09}^{+0.09}$ \\
And~X            & $5.08 \pm 0.03$ & $6.82_{-0.20}^{+0.22}$ & $-2.27 \pm 0.03$ & $-1.00_{-0.29}^{+0.28}$ \\
\cutinhead{MW dSphs}
Fornax           & $7.39 \pm 0.14$ &        $7.87 \pm 0.02$ & $-1.04 \pm 0.01$ & $+0.02_{-0.04}^{+0.04}$ \\
Leo I            & $6.69 \pm 0.13$ &        $7.34 \pm 0.05$ & $-1.45 \pm 0.01$ & $-0.21_{-0.03}^{+0.03}$ \\
Sculptor         & $6.59 \pm 0.21$ &        $7.35 \pm 0.03$ & $-1.68 \pm 0.01$ & $-0.43_{-0.02}^{+0.02}$ \\
Leo II           & $6.07 \pm 0.13$ & $6.86_{-0.06}^{+0.07}$ & $-1.63 \pm 0.01$ & $-0.26_{-0.04}^{+0.04}$ \\
Sextans          & $5.84 \pm 0.20$ & $7.54_{-0.06}^{+0.07}$ & $-1.94 \pm 0.01$ & $-0.56_{-0.08}^{+0.07}$ \\
Ursa Minor       & $5.73 \pm 0.20$ &        $7.75 \pm 0.06$ & $-2.13 \pm 0.01$ & $-0.57_{-0.03}^{+0.03}$ \\
Draco            & $5.51 \pm 0.10$ &        $7.32 \pm 0.06$ & $-1.98 \pm 0.01$ & $-0.44_{-0.05}^{+0.04}$ \\
Canes Venatici I & $5.48 \pm 0.09$ & $7.44_{-0.10}^{+0.13}$ & $-1.91 \pm 0.01$ & $-0.55_{-0.07}^{+0.07}$ \\
\enddata
\tablenotetext{a}{Stellar mass calculated from the product of luminosity and stellar mass-to-light ratio \citep{woo08}.  The luminosities were measured by \citet[][Canes Venatici~I]{martin08} and \citet[][M31 dSphs]{tol12} or otherwise compiled by \citet{mcc12}.}
\tablenotetext{b}{Mass within the half-light radius, taken from Table~\ref{tab:v}.}
\tablecomments{Dynamical masses for MW dSphs were taken from \citet{wol10}.  Abundance information for MW dSphs was drawn from \citetalias{kir13b}.}
\end{deluxetable*}

\begin{figure}
\centering
\includegraphics[width=\linewidth]{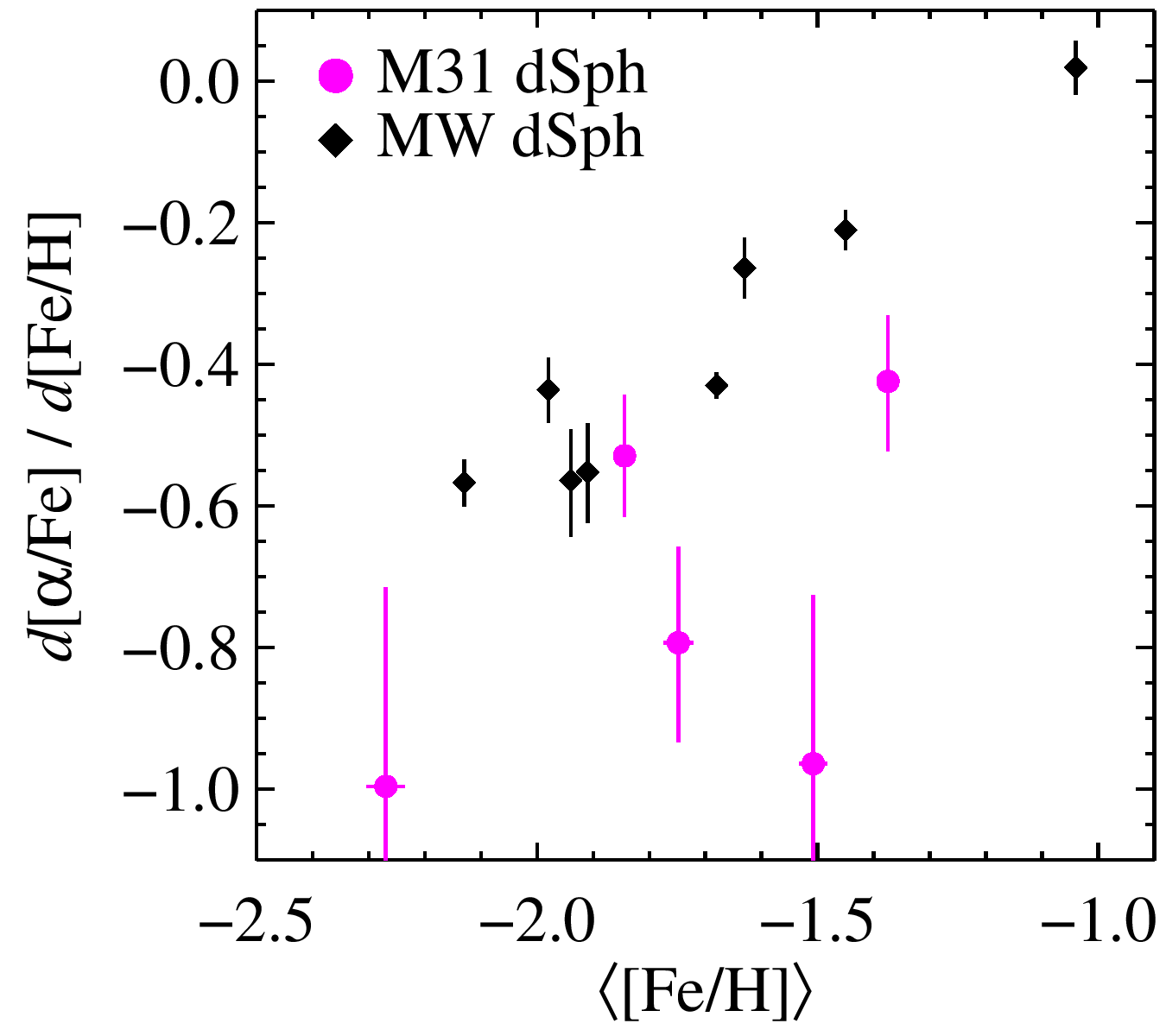}
\caption{The slopes of [$\alpha$/Fe] vs.\ [Fe/H] in M31 dSphs (magenta
  circles) and MW dSphs (black diamonds).\label{fig:alphafeslope}}
\end{figure}

Figure~\ref{fig:alphafe} shows the trends of [$\alpha$/Fe] with
[Fe/H]\@.  The dSphs are listed from top to bottom in order of
decreasing stellar mass.  The MZR is reflected by the leftward
(decreasing [Fe/H]) shift for decreasing $M_*$.  Although the average
metallicities of the stars vary across the dSphs, all of the galaxies
exhibit a decrease of [$\alpha$/Fe] with [Fe/H]\@.

We calculated the slopes of [$\alpha$/Fe] vs.\ [Fe/H] by using maximum
likelihood coupled with a Monte Carlo Markov chain (MCMC)\@.  The
likelihood function parameterizes the line by an angle $\theta$ and a
perpendicular offset $b_{\perp}$.

\begin{equation}
  {\rm [\alpha/Fe]} = {\rm [Fe/H]} \tan \theta + \frac{b_{\perp}}{\cos \theta} \label{eq:line}
\end{equation}
  
\noindent We adopted flat priors on $\theta$ and
$b_{\perp}$.\footnote{This procedure is identical to that of J.\ Wojno
  et al. (in preparation).}  This parameterization avoids the bias
against steep slopes when the line is parameterized by $y = mx + b$
and a flat prior is adopted for $m$ \citep[see][]{hog10}.  The
conversions between the two parameterizations are $m = \tan \theta$
and $b_{\perp} = b \cos \theta$.  We found the best-fit slope with an
MCMC of $10^6$ links following a burn-in period of $10^5$ links.  The
MCMC is based on the Metropolis algorithm.  The slopes $m = d{\rm
  [\alpha/Fe]}/d{\rm [Fe/H]}$ are given in Table~\ref{tab:alphafe} and
also shown in Figure~\ref{fig:alphafeslope}.  The values given are the
$50^{\rm th}$ percentiles of the MCMC chains with 68\% C.L.\ error
bars at the $16^{\rm th}$ and $84^{\rm th}$ percentiles.

A slope of $-1$ would indicate constant [$\alpha$/H], i.e., enrichment
solely by Type~Ia supernovae with no contributions from core collapse
supernovae.  The shallowness of the slope (Table~\ref{tab:alphafe})
for And~VII, the most massive dSph in our sample, indicates that it
maintained a higher average SFR during its star-forming lifetime than
the other dSphs.  The higher SFR allowed it to create a higher ratio
of core collapse to Type~Ia supernovae.  The steeper slopes of the
less massive dSphs reflect inefficient SFHs, wherein Type~Ia
supernovae dominated the chemical evolution.

It could be reasonably expected that the abundance patterns of dwarf
galaxies are more closely related to their dynamical masses than their
stellar masses.  After all, it is the depth of the gravitational
potential well that dictates the galaxy's ability to retain metals
\citep{dek86}.  Table~\ref{tab:alphafe} shows that the average
metallicities and [$\alpha$/Fe] slopes of of M31 dSphs are reasonably
well correlated with dynamical mass.  However, there is essentially no
correlation between dynamical mass and stellar mass or abundance
properties of MW dSphs, as discussed previously in the literature
\citep[e.g.,][]{str08,kir11a}.  One possibility is that the dark
matter in dSphs is much more extended than the stellar distribution.
It is therefore difficult to measure the full depth of the
gravitational potential because there are no luminous tracers of the
mass at large radii.  Unfortunately, our sample size of five M31 dSphs
is not sufficient to make definitive statements about whether the M31
system has a stronger correlation between abundances and dynamical
mass than the MW system.

Figure~\ref{fig:alphafe} also shows MW dSphs with average
metallicities comparable to those of the M31 dSphs.  The MW dSphs are
the same as those shown in Figure~\ref{fig:mwfehhist}.  Like the MDFs,
there is enough diversity in the [$\alpha$/Fe] diagrams among dSphs of
similar $\langle {\rm [Fe/H]} \rangle$ that our ability to draw
quantitative conclusions is limited.  However, there are some
qualitative trends as a function of $M_*$ that persist in both the MW
and M31 samples.  We describe some of these trends below.

Table~\ref{tab:alphafe} shows the $d{\rm [\alpha/Fe]}/d{\rm [Fe/H]}$
slopes of the MW dSphs alongside those of the M31 dSphs.  The slopes
are generally steeper for dSphs of lower $M_*$.  This trend is
consistent with the concept that satellites of both the MW and M31
systems that were able to maintain higher SFRs, as reflected by
shallower [$\alpha$/Fe] slopes, were also able to reach higher stellar
masses \citep[also see][]{tol09,let10,kir11b,lem14}.

Shallower [$\alpha$/Fe] slopes accompany MDFs that are inconsistent
with a simple Leaky Box.  This correlation can be seen by comparing
Table~\ref{tab:gce} to Table~\ref{tab:alphafe}.  More massive dSphs
have narrower MDFs that favor the Pre-Enriched or Accretion Models.
The MDFs suggest that more massive dSphs evolved in the presence of an
external influence, such as gas accretion.  It is possible that these
dSphs were able to maintain higher SFRs for a longer period of time
because they had access to an external gas reservoir.  That could be
why they grew to larger stellar mass.  The tendency for more efficient
star formers (reflected in the [$\alpha$/Fe] slope) to have access to
external gas (reflected in the MDF shape) is present in both the MW
dSphs (\citealt{kir11a}; \citetalias{kir13b}) and the M31 dSphs.

\citet{ski17} measured the SFHs of And~I and And~III, among other M31
satellites, with {\it Hubble Space Telescope} ({\it HST}) imaging
reaching the main sequence turn-off (MSTO)\@.
\citeauthor{ski17}\ found more uniformity in the SFHs among the M31
dSphs than the MW dSphs.  For example, the MW has three dSphs in the
$10^{6-7}~M_{\sun}$ range of stellar mass: Leo~I, Sculptor, and
Leo~II\@.  They have extended, ancient, and intermediate SFHs,
respectively \citep{wei14}.  In contrast, And~I and And~III, which
also fall in that stellar mass range, both have intermediate SFHs
\citep{ski17}.

From shallower {\it HST} imaging, \citet{wei14} measured the SFHs of
And~VII, I, III, and V\@.  The SFHs of And~I and And~III are
consistent with those of \citet{ski17}.  Interestingly,
\citeauthor{wei14}\ found that the SFH of And~VII was as ancient as
that of Sculptor.  The [$\alpha$/Fe] pattern of And~VII is difficult
to reconcile with an ancient SFH\@.  All of the MW dSphs with
exclusively old populations have $\langle {\rm [Fe/H]} \rangle < -1.5$
as well as steeply declining [$\alpha$/Fe].  In contrast, And~VII has
$\langle {\rm [Fe/H]} \rangle = \AndXfehmean \pm \AndXfehmeanerr$, as
well as the shallowest $d{\rm [\alpha/Fe]}/d{\rm [Fe/H]}$ of all the
M31 dSphs in our sample.  And~VII is part of an upcoming, large
Cycle~27 {\it HST} program (GO-15902, PI: D.\ Weisz).  The purpose of
the program is to obtain CMDs of M31 dSphs that reach the MSTO, i.e.,
quality comparable to those obtained by \citet{ski17}.  It will be
interesting to see if And~VII continues to appear exclusively ancient
in these high-quality measurements.

The preference for M31 dSphs to have extended SFHs or at least
intermediate-age populations may extend to its least massive
satellites ($M_* < 10^{5.5}~M_{\sun}$).  \citet{martin17} analyzed
{\it HST}-based CMDs of some of the less massive M31 dSphs, including
And~X\@.  These CMDs reached the horizontal branch (HB) but not the
MSTO\@.  Even the faintest M31 dSphs in their sample
($10^{4.2}~L_{\sun}$) have red HBs, which probably indicate the
presence of intermediate-age populations that are not very metal-poor.
In contrast, low-mass MW dSphs have more prominent blue HBs than red
HBs.  From the perspective of chemical evolution, And~X appears to
have an [$\alpha$/Fe] distribution consistent with MW dSphs of similar
metallicity (Ursa Minor) or slightly larger stellar mass (Canes
Venatici~I), though we measured [$\alpha$/Fe] for only
\AndXnalphafe\ stars in And~X\@.  {\it HST} CMDs show Ursa Minor and
Canes Venatici~I to have slightly older populations than And~I and
And~III \citep{wei14,ski17}.  It is not yet possible to confirm
whether And~X's HB morphology results from a more extended SFH than
its MW counterparts because it does not yet have imaging deep enough
to reach the MSTO\@.  However, And~X will be observed as part of the
aforementioned {\it HST} large program.

\begin{figure}
\centering
\includegraphics[width=\linewidth]{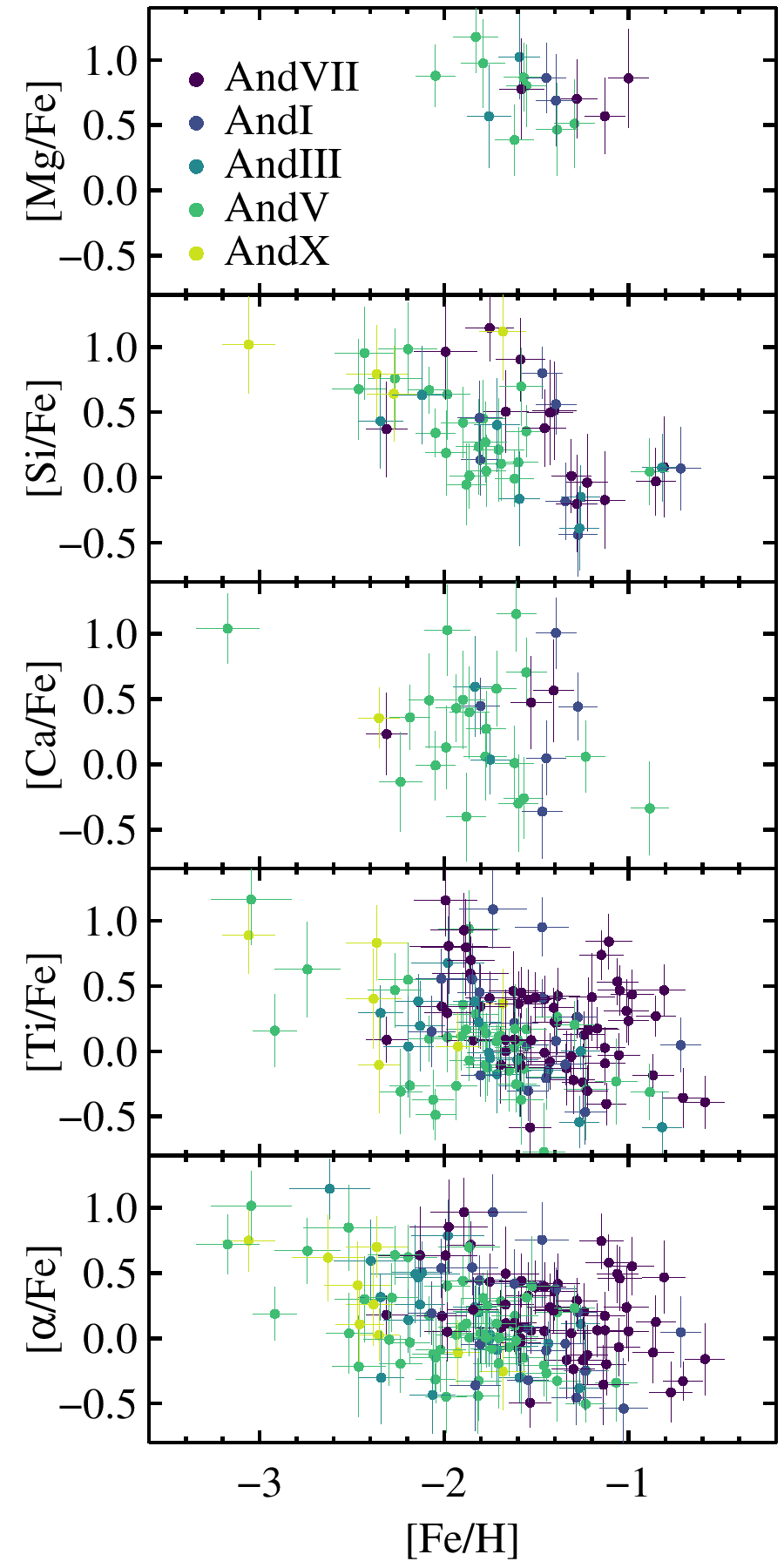}
\caption{Distribution of individual $\alpha$ element ratios (top four
  panels) and [$\alpha$/Fe] (bottom panel; measured from the union of
  Mg, Si, Ca, and Ti absorption lines) as a function of [Fe/H]\@.
  Each M31 dSph is shown with a different color.  Only measurements
  with uncertainties less than 0.4~dex are shown.  This error cut
  imposes a bias against including stars with low values of [Mg/Fe]
  and [Si/Fe] (see Section~\ref{sec:alphafe}).  The legend lists the
  dSphs from largest to smallest stellar mass.\label{fig:mgsicati}}
\end{figure}

Our deep DEIMOS spectroscopy enables us for the first time to measure
individual element ratios in the M31 system.
Figure~\ref{fig:mgsicati} shows [Mg/Fe], [Si/Fe], [Ca/Fe], and [Ti/Fe]
ratios for the five M31 dSphs in our sample.  Each panel shows all
five dSphs to emphasize the difference in [$\alpha$/Fe] distributions
across a range of dSph mass.  The figure excludes measurements with
uncertainties greater than 0.4~dex in either axis.  As a result, there
is a bias against stars with low [Mg/Fe] and [Si/Fe], especially at
low [Fe/H].  The absorption lines of Ca and Ti are generally of lower
excitation potential and thus stronger in red giants \citep[see
  Table~3 of][]{kir10}.  Therefore, the bias against low abundances is
not as strong for [Ca/Fe] and [Ti/Fe].  Appendix~\ref{sec:bias}
further discusses the bias against stars with low [$\alpha$/Fe]
ratios.  The bottom panel of Figure~\ref{fig:mgsicati}, [$\alpha$/Fe],
is not an average of the top four panels.  Instead, it is a
measurement based on the union of Mg, Si, Ca, and Ti absorption lines,
as described at the end of Section~\ref{sec:abundmsmts}.  (This use of
[$\alpha$/Fe] is consistent with the rest of this paper.)  Measuring
[$\alpha$/Fe] is possible at lower metallicity than individual element
ratios because inclusion of a larger number of absorption lines
increases the SNR of the measurement.

The dSphs form a tidy [$\alpha$/Fe]--[Fe/H] sequence in stellar mass.
Although each dSph shows a decline in [$\alpha$/Fe], the average
[$\alpha$/Fe] at fixed [Fe/H] declines with stellar mass.  We have
already seen that the $d{\rm [\alpha/Fe]}/d{\rm [Fe/H]}$ slope
steepens with decreasing $M_*$.  Figure~\ref{fig:mgsicati}
additionally makes clear that [$\alpha$/Fe] begins to decline at lower
[Fe/H] in lower-mass dSphs.  The general appearance is that declining
[$\alpha$/Fe] trends shifts to lower [Fe/H] and becomes steeper as
$M_*$ decreases.

The most straightforward interpretation of these trends is that
low-mass galaxies were susceptible to both heavy gas loss and
inefficient star formation.  The gas loss suppressed the rate at which
[Fe/H] increased, as discussed in relation to the Leaky Box Model in
Section~\ref{sec:feh}.  The low SFRs permitted the rate of Type~Ia
supernovae to overtake the rate of core collapse supernovae.  Larger
galaxies retained more gas and metals, allowing them to reach higher
[Fe/H]\@.  They also had higher SFRs, which allowed them to delay the
diminution of [$\alpha$/Fe] to higher metallicity.

The trends with stellar mass are apparent both in the bulk
[$\alpha$/Fe] ratio as well as individual ratios, especially [Si/Fe]
and [Ti/Fe].  Of these two elements, Si better discriminates between
Type~Ia and core collapse supernovae.  Si is made almost entirely in
core collapse supernovae \citep[i.a.,][]{woo95,kir19}, but Ti is made
in both types of supernovae.  Furthermore, most theoretical
predictions of Ti yields from core collapse supernovae miss the
abundances of metal-poor halo stars by a factor of several
\citep{nom06,kir11b,kob11}.  Nonetheless, Ti has long been observed to
mimic the pattern of ``true'' $\alpha$ elements
\citep[i.a.,][]{ven04}.  Our observations support this observation.
For example, the [Ti/Fe] measurements follow the same trends as the
[Si/Fe] measurements.


\section{M31's Satellites in the Context of Halo Assembly}
\label{sec:discussion}

In this section, we compare the abundance trends of M31's largest
intact dSph\footnote{We are considering that M32 is a compact
  elliptical and NGC~147, NGC~185, and NGC~205 are dwarf ellipticals,
  not dSphs, but the distinction is admittedly arbitrary
  \citep[see][]{mcc12}.}, And~VII, with the GSS and M31's smooth halo.
One purpose of this comparison is to consider the extent to which
galaxies similar to M31's surviving dSphs could have contributed to
its stellar halo.  A similar comparison in the MW system shows that
the inner halo is composed primarily of galaxies that were larger than
the largest surviving dSphs \citep{she01,she03,ven04}.

We use the measurements of [Fe/H] and [$\alpha$/Fe] from previous
papers in this series: \citet{gil19} for the GSS and \citet{esc19a}
for the smooth halo.  We consider only measurements with uncertainties
less than 0.4~dex in both [Fe/H] and [$\alpha$/Fe].  \citet{esc19a}
analyzed four separate fields, called H, S, D, and f130\_2.  The
fields roughly correspond to the halo (at 12~kpc projected from the
center of M31), GSS, outer disk, and a different part of the halo (at
23~kpc).  In detail, each field contains some stars that belong to the
smooth halo and some stars that belong to kinematic substructure.  We
omit stars more likely to belong to substructure than the halo, i.e.,
stars with substructure probability greater than 50\%.  We also omit
stars that have TiO absorption that could distort the abundance
measurements.  \citet{gil19} used the same DEIMOS configuration (1200G
grating) and abundance measurement technique as we used for the dSphs.
\citet{esc19a} used a lower spectral resolution and wider wavelength
range (600ZD grating) and an abundance code adapted to take advantage
of the larger spectral range.  \citet{esc19b} demonstrated the
consistency of the two techniques.

\begin{figure}
\centering
\includegraphics[width=\linewidth]{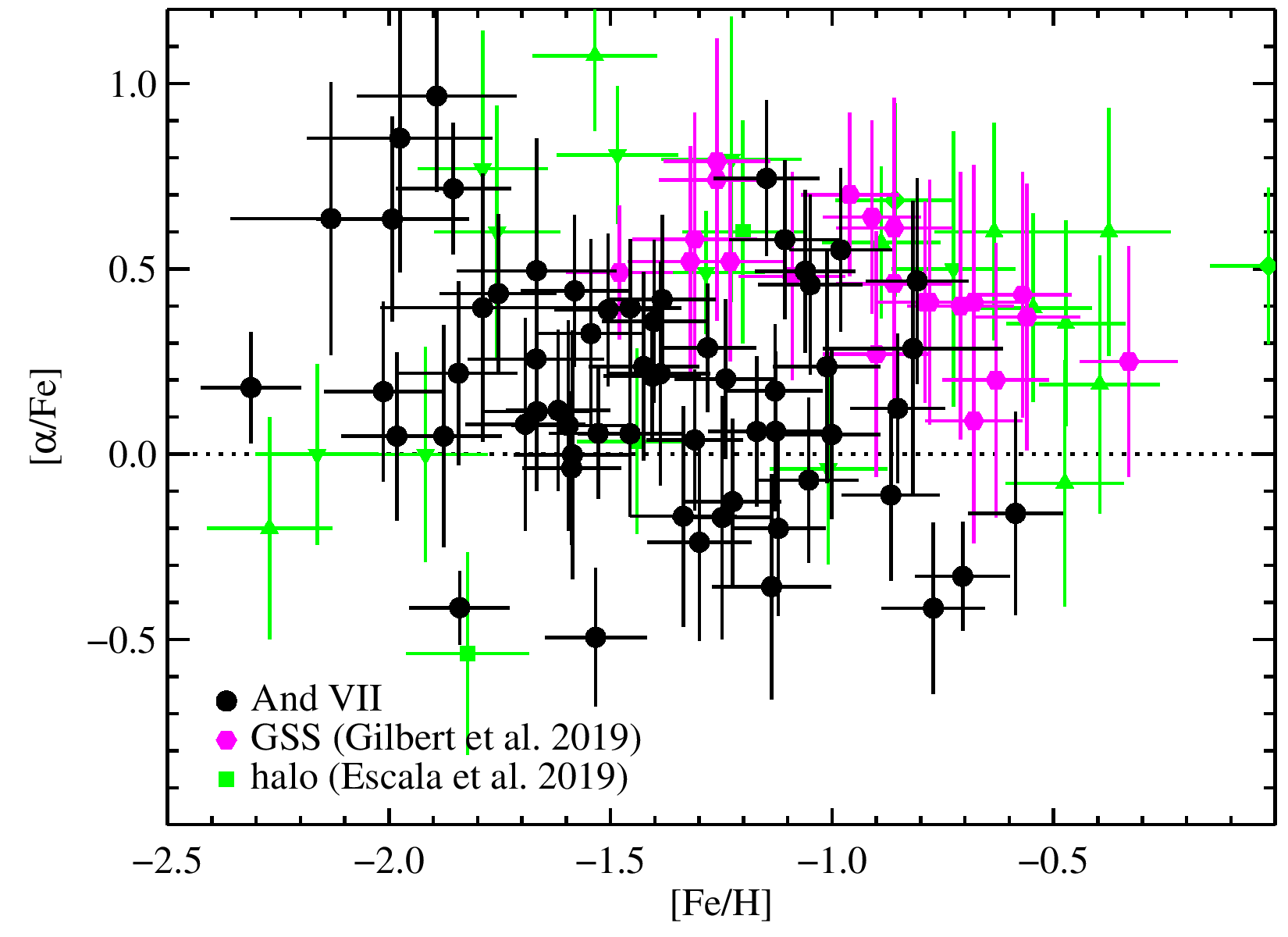}
\caption{Comparison of the abundance trends of And~VII with the GSS
  \citep{gil19} and M31's smooth halo \citep{esc19a}.  Only stars with
  uncertainties less than 0.4~dex in both axes are shown.  The figure
  excludes stars with TiO absorption and halo stars with a probability
  greater than 50\% of belonging to kinematic substructure.  The
  shapes of the halo points identify the fields to which they belong:
  H (squares), S (diamonds), D (upward-pointing triangles), and
  f130\_2 (downward-pointing triangles).  As found by \citet{gil19}
  and \citet{esc19a}, the chemical abundance patterns of M31 dSphs are
  inconsistent with those of the GSS and the metal-rich, inner halo.
  Section~\ref{sec:discussion} interprets the different abundance
  patterns in these systems in the context of the hierarchical
  formation of stellar halos.\label{fig:gss_halo}}
\end{figure}

Figure~\ref{fig:gss_halo} shows the trends of [$\alpha$/Fe] with
[Fe/H] for And~VII, the GSS, and the smooth halo.  It is immediately
clear that each component has a distinct chemical evolution.  And~VII
is more metal-poor on average than the GSS and the halo.  It also has
lower [$\alpha$/Fe] ratios at ${\rm [Fe/H]} \gtrsim -1.5$.  The
chemical distinction from the GSS and halo is even more severe for M31
satellites smaller than And~VII, as shown in
Section~\ref{sec:alphafe}.  We did not show them in
Figure~\ref{fig:gss_halo} for clarity.

This comparison leads us to echo one of the conclusions of
\citet{esc19a}.  Namely, M31's inner halo with ${\rm [Fe/H]} \gtrsim
-1.5$ is not predominantly composed of dwarf galaxies similar to its
surviving dSphs.  The average halo star is more metal-rich and more
$\alpha$-enhanced.  The MZR (Figure~\ref{fig:mzr}) implies that the
typical halo star originated in a galaxy more massive than even
And~VII\@.  The mean metallicity of the halo stars in
Figure~\ref{fig:gss_halo} is $\langle {\rm [Fe/H]} \rangle =
\halofehmean \pm \halofehmeanerr$.  Although these halo stars have a
complicated selection function and could originate from more than one
accretion event, it is clear that the typical halo star is more
metal-rich and $\alpha$-enhanced that the typical star in And~VII\@.
If we assume that the halo stars arose from a single progenitor, the
stellar mass corresponding to its mean metallicity \citepalias{kir13b}
would be $(\halomass \pm \halomasserr) \times 10^7~M_{\sun}$, five
times larger than And~VII\@.  The halo progenitors also experienced
more efficient star formation, which allowed them to maintain high
[$\alpha$/Fe] to a higher metallicity than And~VII\@.

We also restate one of the conclusions of \citet{gil19}: The
progenitor to the GSS was more metal-rich, and therefore more massive,
than M31's surviving dSphs.  \citeauthor{gil19}\ estimated from the
MZR that its original stellar mass was between $5 \times
10^8~M_{\sun}$ and $2 \times 10^9~M_{\sun}$ or possibly higher, given
the sample biases.  In contrast, the stellar mass of And~VII is $1.6
\times 10^7~M_{\sun}$, between those of the MW satellites Fornax and
Leo~I\@.  The decline in [$\alpha$/Fe] also began at [Fe/H] about
0.8~dex higher in the GSS than in And~VII\@.  Therefore, we conclude
that And~VII had a more inefficient SFH, with more gas loss and a
chemical evolution dominated by Type~Ia supernovae, compared to the
GSS\@.

Figure~\ref{fig:gss_halo} is a synopsis of the past, present, and
future of the hierarchical structure formation predicted by
$\Lambda$CDM \citep{rob05,bul05,fon06,joh08}.  In the past, the
nascent M31 accreted a small number of large dwarf galaxies that
experienced high SFRs until the time of accretion.  The stars from
these galaxies have now phase-mixed into the smooth halo.  In the
present, M31 is accreting other dwarf galaxies that likely collapsed
later in the Universe.  The GSS and Sagittarius are two examples of
present-day accretion in M31 and the MW, respectively.  In the future,
the outer halo of M31 will accrete some of its even smaller, intact
satellites that have yet to experience severe tidal disruption.
And~VII and the other dSphs in this study are examples of small,
intact dwarf galaxies that may dissolve into the outer halo in the
distant future.


\section{Summary}
\label{sec:summary}

We presented deep Keck/DEIMOS spectroscopy of five dSph satellites of
M31 spanning a stellar mass range from $1.2 \times 10^5~M_{\sun}$ to
$1.6 \times 10^7~M_{\sun}$.  These dSphs are a subset of those with
previously published shallow spectroscopy \citepalias{var14a}.  The
exposure times (ranging from 4.6 to 11.7~hours) enabled us to achieve
SNR up to an unprecedented \maxsn~\AA$^{-1}$ for individual member
stars.  We measured [Fe/H] and [$\alpha$/Fe] for those stars, with
errors as low as 0.10~dex and 0.12~dex, respectively.  We even
measured the individual abundance ratios [Mg/Fe], [Si/Fe], [Ca/Fe],
and [Ti/Fe] where possible.  We identified \ntot\ member stars across
all five dSphs on the basis of radial velocity, CMD position, and the
strength of the \ion{Na}{1} doublet.  We obtained measurements of
[Fe/H] and [$\alpha$/Fe] for \nfehtot\ and \nalphafetot\ stars,
respectively.

We measured the velocity dispersions and mass-to-light ratios of the
dSphs, and we found them to be consistent with previous measurements
\citep{tol12}.  The majority of the mass in all of the dSphs is dark
matter.  We also found no evidence for galactic rotation.  We placed
upper limits on the rotation velocity between 6 and 10~km~s$^{-1}$.

The M31 dSphs obey the same MZR as MW dSphs and other Local Group
dwarf galaxies.  This conclusion supports past work
(\citetalias{kir13b}; \citetalias{var14a}), but in contrast to that
work, it is based on deep spectroscopy of individual stars.  The M31
dSphs' metallicity distributions have a diversity of shapes.  There is
no strong evidence for a significant difference between the MDF shapes
of M31 and MW dSphs with similar average metallicities.  Like the MW
dSphs, the largest M31 dSphs disfavor a Leaky Box Model of chemical
evolution.  Instead, they require pre-enrichment or gas accretion
during their star-forming lifetimes.

The [$\alpha$/Fe] distributions of the M31 dSphs mimic that of MW
dSphs.  Each dSph has a negative slope of [$\alpha$/Fe] vs.\ [Fe/H],
and that slope is generally steeper for dSphs of lower mass.  This
trend possibly suggests that dSphs of larger stellar mass achieved
their larger masses because they were more efficient at forming stars
and more resilient to gas outflow.  The individual abundance ratios,
especially [Si/Fe], reinforce the conclusions we drew from the bulk
[$\alpha$/Fe] ratios.

The M31 dSphs are more metal-poor and exhibit declining [$\alpha$/Fe]
at lower [Fe/H] than the progenitors of the GSS \citep{gil19} and
M31's smooth halo \citep{esc19a}.  This pattern satisfies the
prediction of $\Lambda$CDM simulations that the inner stellar halos of
MW- or M31-like galaxies are composed of large dwarf galaxies that
were accreted early.  In contrast, smaller galaxies are accreted later
into outer regions of the halo.  These smaller galaxies have lower
metallicities and lower [$\alpha$/Fe] at a given [Fe/H]\@.

\citet{esc19a} measured [$\alpha$/Fe] ratios of M31's smooth halo as
far out as 23~kpc in projected distance from the center of M31.
\citeauthor{esc19a}\ showed tentative evidence that M31's outer halo
\citep[70--140~kpc,][]{var14b} has lower [$\alpha$/Fe] ratios at a
given [Fe/H], in line with the prediction that smaller, less
efficiently star-forming galaxies are accreted into the outer halo.
We predict that additional observations of outer M31 halo stars will
show more dSph-like abundance patterns than the inner halo or GSS\@.
Although it is challenging to observe stars in M31's outer halo due to
their sparseness and the contamination by MW foreground stars, our
series of papers has shown that it is possible to quantify the
chemical evolution in a variety of fields in M31.  Thus, even before
the era of giant telescopes, the M31 system makes possible a test of
$\Lambda$CDM predictions of detailed abundances complementary to the
MW\@.

\acknowledgments

We are grateful to Luis Vargas for providing a data table of M31 dSph
measurements, including effective temperature.  We thank Brent Belland
for helpful discussion on the rotation model and Alexander Ji for
insightful conversation.

This material is based upon work supported by the National Science
Foundation under Grant Nos.\ AST-1614081 and AST-1614569.  ENK
gratefully acknowledges support from a Cottrell Scholar award
administered by the Research Corporation for Science Advancement as
well as funding from generous donors to the California Institute of
Technology.  IE acknowledges support from a National Science
Foundation (NSF) Graduate Research Fellowship under Grant
No.\ DGE-1745301.  PG, SRM, and RLB acknowledge prior funding from
collaborative NSF grants AST-0307842, AST-0307851, AST-0607726,
AST-0807945, AST-1009882, AST-1009973, and AST-1010039.  Support for
this work was provided by NASA through Hubble Fellowship grant
\#51386.01 awarded to RLB by the Space Telescope Science Institute,
which is operated by the Association of Universities for Research in
Astronomy, Inc., for NASA, under contract NAS 5-26555.

We are grateful to the many people who have worked to make the Keck
Telescope and its instruments a reality and to operate and maintain
the Keck Observatory.  The authors wish to extend special thanks to
those of Hawaiian ancestry on whose sacred mountain we are privileged
to be guests.  Without their generous hospitality, none of the
observations presented herein would have been possible.  We express
our deep gratitude to the staff at academic and telescope facilities
whose labor maintains spaces for scientific inquiry.

\facility{Keck:II (DEIMOS)}
\software{spec2d \citep{coo12,new13}, MOOG \citep{sne73,sne12}, ATLAS9
  \citep{kur93}, MPFIT \citep{mar12}}

\bibliography{ms}
\bibliographystyle{apj}


\appendix
\section{Comparison to V14a}
\label{sec:vargas}

\begin{figure*}
\centering
\includegraphics[width=0.85\linewidth]{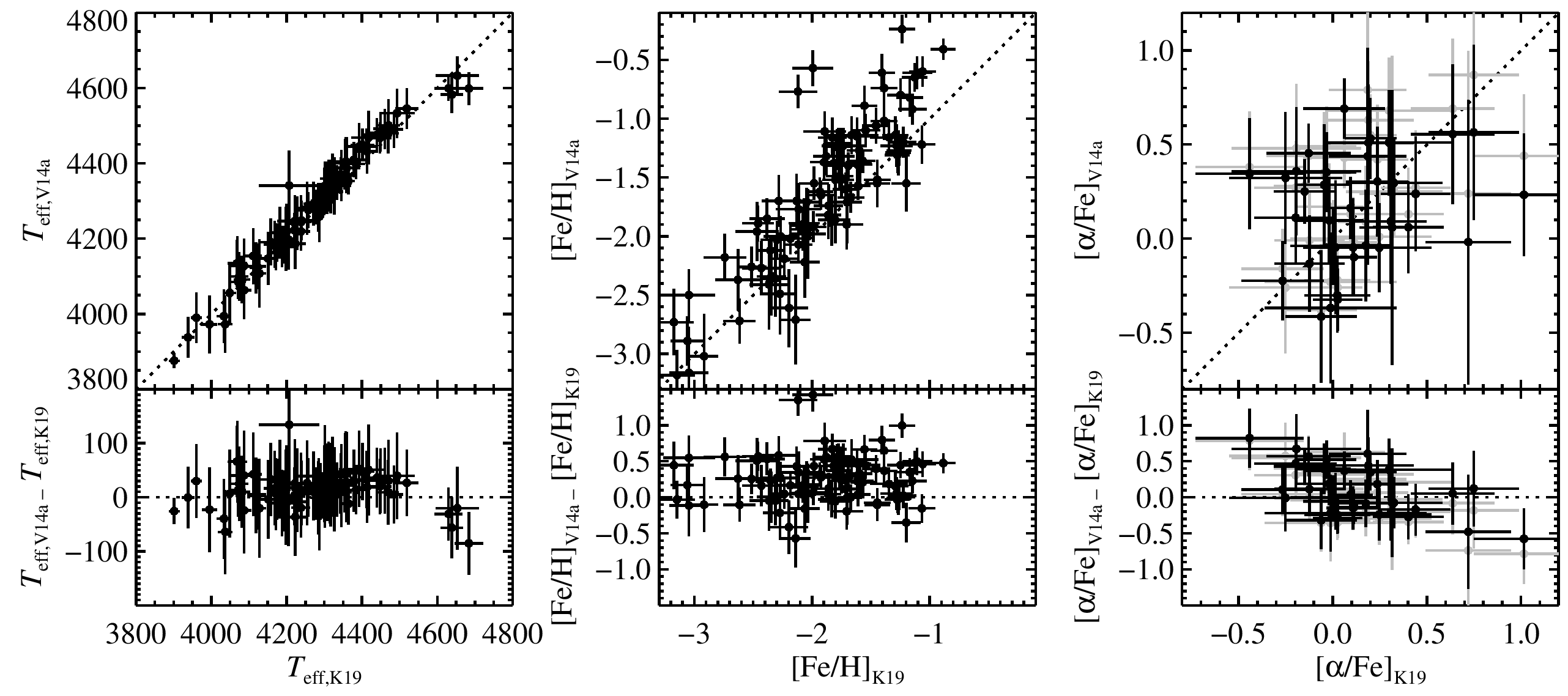}
\caption{Comparison of our measurements (``K19'') of $T_{\rm eff}$,
  [Fe/H], and [$\alpha$/Fe] with those of \citetalias{var14a}.  The
  gray points in the right panel show the ``corrected'' values of
  ${\rm [\alpha/Fe]}_{\rm V14a}$, as described in
  Appendix~\ref{sec:vargas}.\label{fig:vargas}}
\end{figure*}

\citetalias{var14a} measured [Fe/H] and [$\alpha$/Fe] for many of the
stars in our sample.  The advantage of our study is longer exposures,
which resulted in spectra with higher SNR\@.  This appendix compares
our measurements to those of \citetalias{var14a}.
Figure~\ref{fig:vargas} summarizes the comparison.  To be consistent
with our sample selection, we restrict both samples to measurements
with uncertainties less than 0.4~dex.  \citetalias{var14a} provided a
table of abundances.  We obtained a table including their measurements
of $T_{\rm eff}$ from L.~Vargas (private communication).

\citetalias{var14a}'s measurements of [Fe/H] are significantly higher
than ours.  The median difference, ${\rm [Fe/H]}_{\rm V14a}-{\rm
  [Fe/H]}_{\rm K19}$, is $\medfehdiff$.  The mean difference, weighted
by the inverse square of the quadrature sum of the uncertainties, is
$\meanfehdiff \pm \meanfehdifferr$.  We cannot readily explain this
systematic offset because our two studies used nearly identical
techniques to measure abundances.  The most significant difference was
the set of isochrones used to determine the initial photometric
$T_{\rm eff}$.  If this is the source of the difference in [Fe/H],
there should be an accompanying difference in $T_{\rm eff}$ in the
sense that \citetalias{var14a}'s measurements of $T_{\rm eff}$ should
be larger than ours by roughly 300~K \citep[see Table~6 of][]{kir10}.
Figure~\ref{fig:vargas} shows that there is no such offset.

The standard deviation of the differences is $\stddevfehdiff \pm
\stddevfehdifferr$.  We also computed the standard deviation of the
metallicity difference normalized by measurement uncertainty.

\begin{equation}
{\rm stddev}\left( \frac{{\rm [Fe/H]}_{\rm V14a} - {\rm [Fe/H]}_{\rm K19}}{\sqrt{\delta{\rm [Fe/H]}_{\rm V14a}^2 + \delta{\rm [Fe/H]}_{\rm K19}^2}} \right) = \stddevfehdiffnorm \pm \stddevfehdiffnormerr \label{eq:stddevfeh}
\end{equation}

\noindent This quantity would be 1.0 if the measurement uncertainty
completely explained the scatter.  The fact that it is larger than 1.0
indicates that there is an additional source of discrepancy.  Two
outliers serve to increase the scatter.  When the two points with
${\rm [Fe/H]}_{\rm V14a} - {\rm [Fe/H]}_{\rm K19} > +1.0$ are
excluded, the quantity in Equation~\ref{eq:stddevfeh} drops to
$\stddevfehdiffnormcut \pm \stddevfehdiffnormcuterr$.

\citetalias{var14a} measured [$\alpha$/Fe] in a manner similar to us.
However, they also computed a correction to [$\alpha$/Fe].  The
correction was intended to bring [$\alpha$/Fe] into closer agreement
with the arithmetic average of [Mg/Fe], [Si/Fe], [Ca/Fe], and [Ti/Fe].
We reversed this correction so that we could compare like quantities.
Nonetheless, we still present \citetalias{var14a}'s corrected values
as gray points in the right panel of Figure~\ref{fig:vargas}.

Unlike [Fe/H], there is no significant offset in [$\alpha$/Fe] between
our two samples.  The median difference, ${\rm [\alpha/Fe]}_{\rm
  V14a}-{\rm [\alpha/Fe]}_{\rm K19}$, is $\medalphafediff$.  The mean
difference, weighted by the inverse square of the quadrature sum of
the uncertainties, is $\meanalphafediff \pm \meanalphafedifferr$.  The
standard deviation of the differences is $\stddevalphafediff \pm
\stddevalphafedifferr$.  When the difference is normalized by the
measurement uncertainties, as in Equation~\ref{eq:stddevfeh}, the
standard deviation is $\stddevalphafediffnorm \pm
\stddevalphafediffnormerr$.  We conclude that the measurement
uncertainties account for the scatter in [$\alpha$/Fe].

\section{Bias Against Stars with Low [$\alpha$/Fe]}
\label{sec:bias}

\begin{figure}
\centering
\includegraphics[width=0.5\linewidth]{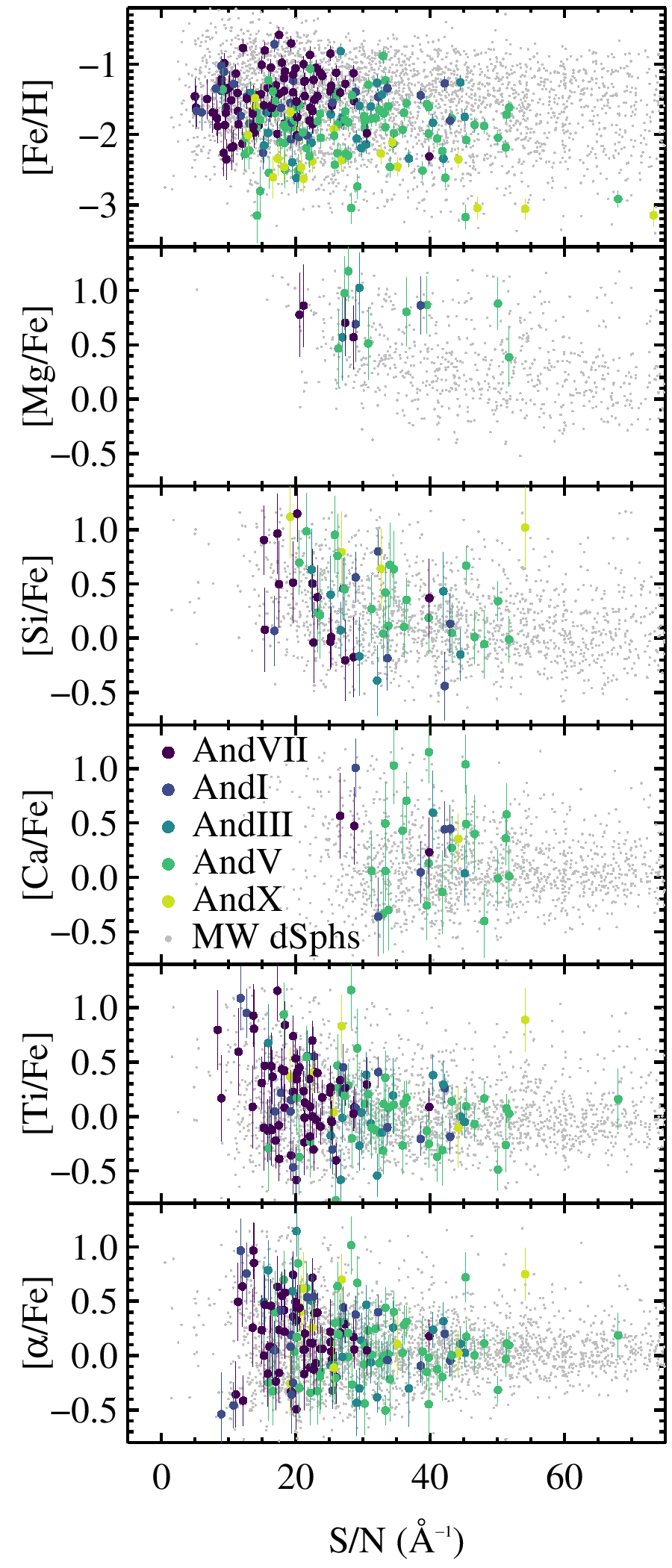}
\caption{Metallicity ([Fe/H], top panel), individual $\alpha$ element
  ratios (top four panels) and [$\alpha$/Fe] (bottom panel) as a
  function of SNR\@.  The data for the M31 dSphs are the same as shown
  in Figure~\ref{fig:mgsicati}.  This figure also includes
  measurements in MW dSphs \citep{kir10}.  Only measurements with
  uncertainties less than 0.4~dex are shown.\label{fig:bias}}
\end{figure}

In Section~\ref{sec:alphafe}, we presented the distribution of
individual $\alpha$ element ratios with respect to iron
(Figure~\ref{fig:mgsicati}).  There was an apparent bias against stars
with low abundance ratios, especially apparent in [Mg/Fe] and [Si/Fe].
We further investigate this bias with Figure~\ref{fig:bias}, which
shows the abundance ratios vs.\ SNR\@.  The figure also includes
measurements in MW dSphs.

There is an increasing dearth of stars with low [Mg/Fe] and [Si/Fe] as
the SNR decreases.  The Mg and Si lines used in our spectral synthesis
are the weakest among the elements we measured because they generally
originate from electron levels with high excitation potentials ($\sim
5$~eV\@).  As a result, they are the first to become undetectable as
the SNR and/or abundance drops.  Our cut on uncertainty of 0.4~dex
further magnifies the bias, which presents as the absence of stars in
a wedge in the lower left of the top two panels in
Figure~\ref{fig:bias}.

The Ca and Ti lines are stronger and therefore still usable at low
SNR\@.  The effect of low SNR is to cause the abundance ratios to fan
out rather than to create a wedge that lacks stars.  Still, there is
some asymmetry in the distribution of abundance measurements at low
SNR\@.  Because it is easier to measure high abundances than low
abundances, the points tend to scatter high, again leading to a bias
against low abundance ratios.

The average [$\alpha$/Fe] is based on the union of all the absorption
lines of Mg, Si, Ca, and Ti.  It is not the average of the abundance
measurements, but it is a separate measurement that uses more
information than is possible when isolating a single element.  As a
result, the average [$\alpha$/Fe] is less biased at low SNR than the
individual $\alpha$ element measurements.

This analysis applies to abundance measurements from spectra obtained
with DEIMOS's 1200G diffraction grating.  The 600ZD grating accesses a
wider wavelength range, including strong Mg lines.  Therefore, the
biases shown here would not necessarily apply to 600ZD measurements
\citep[e.g.,][]{esc19a,esc19b}.

\end{document}